\documentclass[11pt,a4paper]{article}

\usepackage{jheppub}
\pdfoutput=1                     

\usepackage[T1]{fontenc} 
\usepackage[utf8]{inputenc} 
\usepackage{graphicx}
\graphicspath{{./Figs/}}
\usepackage{slashed}
\usepackage{mathtools}        
\usepackage{bbm}              
\usepackage{bm}               
\usepackage{amsmath, amssymb}
\usepackage{bbold}
\usepackage{tikz} 
\usepackage{feynmp-auto}          
\usetikzlibrary{positioning,arrows.meta,decorations.markings,decorations.pathmorphing}
\usepackage{upgreek} 
\usepackage[titletoc]{appendix}

\usepackage{cleveref} 
    \crefname{equation}{}{}
    \crefname{figure}{}{}    
    \crefname{table}{}{}
    \crefname{section}{}{}   
    \crefname{appendix}{}{}
    \crefname{footnote}{}{}
    
    \crefalias{subequation}{equation}

\newcommand{\sfrac}[2]{{\textstyle{\frac{#1}{#2}}}} 
\newcommand{\idx}[1]{{\rm \scriptscriptstyle{(#1)}}}

\newcommand{\lRidx}[1]{{\lambda^{\mkern-3mu\idx{#1}}_{\rm R}}}
\newcommand{\widebar}[1]{\mkern 1.3mu\overline{\mkern-1.3mu#1\mkern-1.3mu}\mkern 1.3mu}
\newcommand{\lt}{{\scriptscriptstyle <}} 
\newcommand{\overbar}[1]{\mkern 1.5mu\overline{\mkern-1.5mu#1\mkern-1.5mu}\mkern 1.5mu} 

\def\phR{{\varphi_{\rm R}}}
\def\G{{\Delta}}
\def\GR{{\Delta}_{\rm R}}

\title{Non-equilibrium dynamics of a scalar field with quantum backreaction}

\author[a,b]{Kimmo Kainulainen}
\author[a,b]{and Olli Koskivaara}

\affiliation[a]{Department of Physics, University of Jyväskylä,\\ P.O.~Box 35 (YFL), FI-40014 Jyväskylä, Finland}
\affiliation[b]{Helsinki Institute of Physics, University of Helsinki,\\ P.O.~Box 64, FI-00014 Helsinki, Finland.}

\emailAdd{kimmo.kainulainen@jyu.fi}
\emailAdd{olli.a.koskivaara@student.jyu.fi}

\abstract{We study the dynamical evolution of coupled one- and two-point functions of a scalar field in the  
2PI framework at the Hartree approximation, including backreaction from out-of-equilibrium modes. We renormalize the 2PI equations of motion in an on-shell scheme in terms of physical parameters. We present the Hartree-resummed renormalized effective potential at finite temperature and critically discuss the role of the effective potential in a non-equilibrium system. We follow the decay and thermalization of a scalar field from an initial cold state with all energy stored in the potential, into a fully thermalized system with a finite temperature. We identify the non-perturbative processes of parametric resonance and spinodal instability taking place during the reheating stage. In particular we study the unstable modes in the region where the vacuum 1PI effective action becomes complex and show that such spinodal modes can have a dramatic effect on the evolution of the one-point function. Our methods can be easily adapted to simulate reheating at the end of inflation.}

\keywords{Thermal Field Theory, Quantum Dissipative Systems, Nonperturbative Effects}

\arxivnumber{2105.09598}

\begin{document}
\maketitle



%
\section{Introduction}
\label{sec:intro}
%

Classical scalar fields coupled to quantum matter play an important role in various settings in cosmology. They are used to study the creation of seed perturbations for structure formation, reheating processes, particle production and the creation of baryon asymmetry. Almost exclusively in these treatments it is assumed that the scalar field evolves in some classical, possibly quantum corrected but fixed effective potential. One rarely accounts for the backreaction of the non-equilibrium quanta that may be created during the dynamical process. However, such quanta may be produced copiously during out-of-equilibrium phase transitions~\cite{Traschen:1990sw,Amin:2014eta} by parametric resonance~\cite{Kofman:1994rk,kofman:1997yn,Greene:1997fu,Braden:2010wd,PhysRevLett.91.111601} or by spinodal instability~\cite{Calzetta:1989bj,Guth:1985ya,Weinberg:1987vp,Braden:2010wd,Dufaux:2006ee,Fairbairn:2018bsw,Markkanen:2015xuw}, and they could significantly affect the evolution of the system~\cite{Boyanovsky:1992vi,Boyanovsky:1993pf,PhysRevD.65.065019,Arrizabalaga:2004iw,Arrizabalaga:2005tf}. In this paper we study the effects of quantum backreaction on the scalar field evolution using two-particle irreducible (2PI) effective action methods. 

A crucial step in the rigorous analysis of the problem is performing a consistent renormalization of the equations of motion derived from the 2PI effective action. This is a highly non-trivial task, because in any finite truncation of the 2PI expansion, a number of auxiliary vertex and self-energy functions appear that require setting up consistent renormalization conditions~\cite{Berges:2005hc}. Other works on the renormalization of 2PI-truncated theories include for example references~\cite{PhysRevD.65.025010,PhysRevD.65.105005,PhysRevD.83.125026}. In this paper we carefully go through the renormalization of our model using the method of cancellation of the sub-divergences~\cite{Fejos:2007ec,Arai:2012sh,Pilaftsis:2013xna,Pilaftsis:2017enx}. We emphasize that while the renormalization counterterms are constants, the divergences that get subtracted, and hence also the vacuum state of the system, depend on the infrared physics, such as temperature, or even the shape of the non-equilibrium particle spectrum. 

To be specific, we study a simple $\lambda \phi^4$-model with a spontaneous symmetry breaking tree-level potential. We work in the Hartree approximation and perform the auxiliary renormalizations using the $\overbar {\mathrm{ MS}}$ subtraction scheme. The renormalized equations of motion and the 2PI effective action are however scale independent and completely specified in terms of physical parameters. We present explicit results for the vacuum and finite temperature effective potentials as well as for the vacuum potential in the presence of non-equilibrium fluctuations. We stress that in the non-equilibrium case the effective potential can only be constructed \emph{a posteriori} and it is not in general a useful quantity for solving the equations of motion.

With our renormalized equations we can follow in real time how the potential energy of the classical field is transferred into quantum fluctuations by the non-perturbative processes. We identify a strong parametric resonance, even though our self-coupled system is too complicated to admit a comprehensive analytical stability analysis. We also show that due to backreaction from spinodal instability the field can pass through a potential barrier even when starting with less energy than the initial barrier height. We also follow the full thermal history of a system that starts with pure potential energy, until it is fully thermalized with nearly all of its energy stored in thermal plasma. We also show that at the initial stages of reheating the quantum system is highly coherent, but the coherence is gradually erased by interactions as the system thermalizes.  

This paper is organized as follows. In section~\cref{sec:2PI} we review the 2PI effective action techniques and introduce our truncation scheme, the Hartree approximation. In section~\cref{sec:renormalization-main} we show how to self-consistently renormalize the 2PI equations of motion and express them in terms of physical quantities. We also study both resummed vacuum and thermal effective potentials in the Hartree case and compare them with other approximations. In section~\cref{sec:wigner} we write our equations of motion in the Wigner space in terms of moment functions following references~\cite{Herranen:2008di,Herranen:2010mh}, and also complement the equations with phenomenological friction terms. Section~\cref{sec:results} is dedicated to numerical results. We compute the evolution of various quantities, such as the classical field, particle number and coherence functions using the fully coupled 2PI equations. Finally, section~\cref{sec:conc} contains our conclusions.

%
\begin{figure}[t]
\begin{center}
\includegraphics[scale=0.95]{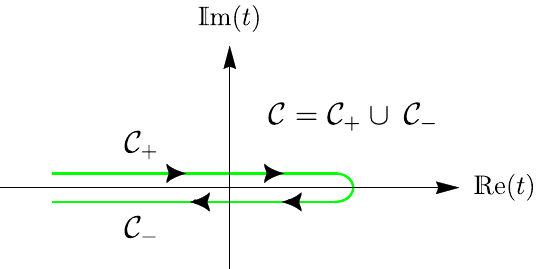}
\end{center}
\caption{The Keldysh contour in the complex time plane, running from some initial time to an arbitrary future time and back again.}
\label{fig:ctp}
\end{figure}
%

%
\section{2PI effective action and equations of motion}
\label{sec:2PI}
%

We will study the non-equilibrium dynamics of a scalar field theory with the potential $V(\phi) = -\frac{1}{2}\mu^2\phi^2 + \frac{1}{4!}\lambda\phi^4$ using the two-particle irreducible (2PI) effective action technique of non-equilibrium quantum field theory~\cite{Cornwall:1974vz,Berges:2004yj}. The 2PI effective action for this theory is
\begin{equation}
\Gamma_{\rm 2PI} [\varphi, \G] = \mathcal{S}[\varphi] - \frac{\mathrm{i}}{2}\mathrm{Tr}_{\mathcal{C}}\bigl[ \ln (\G)\bigr] + \frac{\mathrm{i}}{2}\mathrm{Tr}_{\mathcal{C}}\bigl[\G_0^{-1}\G\bigr] + \Gamma_2[\varphi, \G],
\label{2PI-effective-action}
\end{equation}
where $\varphi(x)$ is the classical field and $\G(x,y)$ is the classical connected two-point function
and the trace contains integration over the Keldysh contour~\cite{Keldysh:1964ud} $\mathcal{C}$ of figure \ref{fig:ctp}. Moving to a real-time representation the classical action can be written as $\mathcal{S}[\varphi] = \sum_{a=\pm } a\delta^{ab}\mathcal{S}[\varphi_b]$, where $a$ and $b$ indicate the branch on the complex time-contour, and
\begin{equation}
\mathcal{S}[\varphi_b] = \int {\rm d}^4x \bigg[ \frac{1}{2}(\partial_\mu\varphi_b)^2  + \frac{1}{2}\mu^2\varphi_b^2 - \frac{1}{4!}\lambda \varphi_b^4 \bigg].
\label{eq:classical-effective-action}
\end{equation}
Similarly, the inverse classical propagator is given by
\begin{equation}
\mathrm{i}\G_{0,ab}^{-1}(x,y;\varphi) = - \bigg( \Box_x - \mu^2 + \frac{1}{2}\lambda \varphi_a^2\bigg)\delta^{(4)}(x-y)\delta_{ab}.
\label{eq:free-inverse-propagator}
\end{equation}
Finally, $\Gamma_2$ consists of all 2PI vacuum graphs with lines corresponding to the full propagator $\G$ and interactions inferred from the shifted action
\begin{equation}
S_{\rm int}\bigl[\varphi,\phi_q\bigr] = - \sum_{a=\pm} a\delta^{ab} \int {\rm d}^4x\bigg(\frac{1}{3!}\lambda\varphi_b \phi_{qb}^3 + \frac{1}{4!}\lambda\phi_{qb}^4\bigg),
\label{eq:shifted-action}
\end{equation}
where $\phi = \varphi + \phi_q$ and $\phi_q$ is the quantum field.

The stationarity conditions of $\Gamma_{\rm 2PI}$ will give the equations of motion for the one- and two-point functions: 
\begin{equation}
\frac{\delta \Gamma_{\rm 2PI}}{\delta \varphi_a}=0
\qquad {\rm and} \qquad
\frac{\delta \Gamma_{\rm 2PI}}{\delta \Delta_{ab}}=0.
\label{eq:stationarity}
\end{equation}
When the classical solution to the latter equation, parametrized in terms of $\varphi$, is reinserted back into the effective action, we \emph{formally} recover the 1PI action $\hat \Gamma_{\rm 1PI}[\varphi ] = \Gamma_{\rm 2PI}[\varphi,\Delta[\varphi]]$. In the full dynamical case the two equations are however strongly coupled and should be solved simultaneously, as we will do in our study. For the classical field $\varphi_+(x) = \varphi_-(x)$ and we may drop the branch index and find:
\begin{equation}
\bigg[ \Box_x - \mu^2 + \frac{1}{6}\lambda\varphi^2(x) + \frac{1}{2} \lambda\G(x,x)\bigg] \varphi(x) = \frac{\delta \Gamma_2}{\delta \varphi(x)}.
\label{eq:eom-for-one-point-function}
\end{equation}
We also left the branch indices out from the local correlation function $\G(x,x)$, which is the same for all components of the two-point function $\G^{ab}(x,y)$. The stationarity condition for $\G^{ab}(x,y)$ leads to the Schwinger--Dyson equation 
\begin{equation}
\Big[ \Box_x - \mu^2 + \sfrac{1}{2}\lambda \varphi^2(x)\Big]\mathrm{i}\G^{ac}(x,y) = a\delta^{ac} 
\delta^{(4)}(x-y) + b\hspace{-.2em}\int \hspace{-.2em} \mathrm{d}^4z \,  \Pi^{ab}(x,z)\G^{bc}(z,y),
\label{eq:eom-for-two-point-function}
\end{equation}
where summation over $b$ is implied and the self-energy function is given by
\begin{equation}
  \Pi^{ab}(x,y) = 2\mathrm{i}ab \, \frac{\delta \Gamma_2[\varphi, \G]}{\delta \G^{ba}(y,x)} 
  = a\delta^{ab}\delta^{(4)}(x-y)\Pi_{\mathrm{sg}}(x) + \Pi_{\mathrm{nsg}}^{ab}(x,y).
\end{equation}
To proceed we also have to specify an approximation for the interaction term $\Gamma_2$. 

%
\begin{figure}[t]
\begin{fmffile}{diagram}
\begin{equation*}
\hspace{-1em}
\parbox{17mm}
{
\begin{fmfgraph*}(70,40)
\fmfleft{i}
\fmfright{o}
\fmf{phantom}{i,v,v,o}
\fmf{plain}{v,v}
\fmf{plain,left=90}{v,v}
\fmfv{label=\scriptsize{$\sim\!\lRidx{0}\!+\delta_\lambda^\idx{0}$},label.angle=-90,label.dist=.5w}{v}
\end{fmfgraph*}
}
\hspace{0.7em}
+
\hspace{1.3em}
\parbox{20mm}
{
\begin{fmfgraph*}(40,40)
\fmfleft{v1}
\fmf{plain,left,tension=.3}{v1,v2,v1}
\fmf{plain}{v1,v2}
\fmfright{v2}
\fmf{phantom}{v1,v3,v2}
\fmfv{label=\scriptsize{$\sim\!\left(\lRidx{1}\!+\delta_\lambda^\idx{1}\right)^2$},label.angle=-90,label.dist=.8w}{v3}
\end{fmfgraph*}
}
\hspace{-0.2em}
+
\hspace{1.4em}
\parbox{20mm}
{
\begin{fmfgraph*}(40,40)
\fmfleft{v1}
\fmf{plain,left,tension=.3}{v1,v2,v1}
\fmf{plain,left=0.4}{v1,v2,v1}
\fmfright{v2}
\fmf{phantom}{v1,v3,v2}
\fmfv{label=\scriptsize{$\sim\!\left(\lRidx{0}\!+\delta_\lambda^\idx{0}\right)^2$},label.angle=-90,label.dist=.8w}{v3}
\end{fmfgraph*}
}
\hspace{-0.2em}
+
\hspace{1.4em}
\parbox{20mm}
{
\begin{fmfgraph*}(40,40)
\fmfforce{(0w,0.5h)}{v1}
\fmfforce{(1.0w,0.5h)}{v2}
\fmfforce{(.5w,1h)}{v3}
\fmfforce{(0.07w,0.25h)}{v4}
\fmfforce{(0.93w,0.25h)}{v5}
\fmfforce{(0.5w,0.5h)}{v6}
\fmf{plain,left,tension=.3}{v1,v2,v1}
\fmf{plain}{v6,v3}
\fmf{plain}{v6,v4}
\fmf{plain}{v6,v5}
\fmfv{label=\scriptsize{$\sim\!\left(\lRidx{1}\!+\delta_\lambda^\idx{1}\right)^4$},label.angle=-90,label.dist=.8w}{v6}
\end{fmfgraph*}
}
\hspace{-0.2em}
+
\hspace{1.4em}
\parbox{20mm}
{
\begin{fmfgraph*}(40,40)
\fmfforce{(0w,0.5h)}{v1}
\fmfforce{(1.0w,0.5h)}{v2}
\fmfforce{(.5w,0h)}{v3}
\fmfforce{(0.2w,0.9h)}{v4}
\fmfforce{(0.8w,0.9h)}{v5}
\fmf{plain,left,tension=.3}{v1,v2,v1}
\fmf{plain}{v3,v4}
\fmf{plain}{v3,v5}
\fmfforce{(0.5w,0.5h)}{v6}
\fmfv{label=\scriptsize{$\sim\!\left(\lRidx{0}\!+\delta_\lambda^\idx{0}\right)\left(\lRidx{1}\!+\delta_\lambda^\idx{1}\right)^2$},label.angle=-90,label.dist=.8w}{v6}
\end{fmfgraph*}
}
\hspace{-0.2em} +\hspace{0.4em} \cdots
\end{equation*}
\end{fmffile}
\vspace{2.5em}
\caption{The first few terms contributing to $\Gamma_2$, including their precise coupling constant dependences.}
\label{fig:gamma2-expansion}
\end{figure}
%

%
\subsection{Hartree approximation}
\label{sec:hartree}
%

The first few terms contributing to $\Gamma_2$, arising from the action~\cref{eq:shifted-action}, are shown in figure~\cref{fig:gamma2-expansion} (the role of the indices in the couplings is related to renormalization and will be explained in the next section). In this work we shall work in the Hartree approximation, which includes only the first term in the series, given by
\begin{equation}
  \Gamma^{\rm H}_2 = -\frac{\lambda}{8} \int \mathrm{d}^4x \, \G^2(x,x).
\end{equation}
In this case the self-energy has only a singular or local part:
\begin{equation}
  \Pi_{\mathrm{sg}}(x) = -\frac{\mathrm{i}\lambda}{2} \G(x,x), 
\end{equation}
while $\Pi_{\mathrm{nsg}}^{ab}(x,y) = 0$. Obviously $\partial \Gamma^{\rm H}_2/\partial\varphi =0$ as well, so there is no contribution to equation~\cref{eq:eom-for-one-point-function} in the Hartree approximation. We can now write the non-renormalized equations of motion compactly as
\begin{subequations}
\begin{align}
\biggl[ \Box_x - \mu^2 + \frac{1}{6}\lambda\varphi^2(x) + \frac{1}{2} \lambda\G(x,x)\biggr] \varphi(x) &= 0 ,
\label{eq:eom-for-1pt-fun}
\\
\biggl[ \Box_x - \mu^2 + \frac{1}{2}\lambda \varphi^2(x) + \frac{1}{2} \lambda\G(x,x) \biggr]\mathrm{i}\G^{ab}(x,y) &= a\delta^{ab} \delta^{(4)}(x-y),
\label{eq:eom-for-2pt-fun}
\end{align}
\end{subequations}
Eventually we will move to the Wigner space defined in section~\cref{sec:wigner} and solve these equations numerically in some example cases for homogeneous systems, but before we can do that, we have to address the divergences in $\G^{ab}$ and in particular in the local correlation function $\G(x,x)$.

%
\section{Renormalization}
\label{sec:renormalization-main}
%

Systematic renormalization in the context of the 2PI expansion was thoroughly discussed in reference~\cite{Berges:2005hc}. Here we use the method introduced in reference~\cite{Fejos:2007ec}, and later used in references~\cite{Arai:2012sh,Pilaftsis:2017enx}, and we include also a connection to physical parameters. The key issue is that any finite order truncation of $\Gamma_2[\varphi,\G]$ leads to an approximation for $\hat \Gamma_{\rm 1PI}[\varphi]$ that contains infinite resummations of 1PI diagrams and the associated counterterms. This gives rise to a number of {\em auxiliary} $n$-point functions which need independent renormalization conditions. These conditions can be defined by requiring that all sub-divergences cancel~\cite{Fejos:2007ec}, but one needs to introduce a different renormalized parameter for each different operator. To be precise, all $n$-point functions can be classified in terms of the number of classical fields that connect to them, and all functions that are connected also to propagator lines are auxiliary.

Below we shall first renormalize the auxiliary $n$-point functions in the $\widebar{\rm MS}$-scheme and show that the resulting 1PI action is independent of the renormalization scale. We start by defining the renormalized fields, propagators, couplings and masses:
\begin{equation}
\begin{aligned}
  \phi &\equiv Z^{1/2}_\idx{2} \phi_{\rm R}, &\hspace{5em}  
\lambda &\equiv \lRidx{I} + \delta\lambda^\idx{I}, \\
  \G &\equiv Z_\idx{0} \G_{\rm R},    &\hspace{5em}
 \mu^2 &\equiv \mu^2_{{\rm R}\idx{I}} - \delta\mu^2_\idx{I},
\end{aligned}
\label{eq:ren-cond}
\end{equation}
where the index, ${\rm I}=0,1,2,4$, follows the power of the classical field associated with the $n$-point function. Written in terms of the renormalized quantities, the 2PI effective action becomes:
\begin{equation}
\begin{split}
\Gamma_{\rm 2PI} [\phR, \GR] &= \mathcal{S}[\phR] - \frac{\mathrm{i}}{2}\mathrm{Tr}_{\mathcal{C}}\bigl[\ln( Z_\idx{0}\GR) \bigr] + \frac{\mathrm{i}}{2}\mathrm{Tr}_{\mathcal{C}}\left[\G_{0\rm R}^{-1}\GR\right] 
 \\
&+ \delta\mathcal{S}[\phR] 
+ \frac{\mathrm{i}}{2}\mathrm{Tr}_{\mathcal{C}}\bigl[\delta \G_{0}^{-1}\GR\bigr] 
+  \Gamma_2\Bigl[\phR, \GR;\lRidx{I} + \delta_\lambda^\idx{I}\Bigr],
\label{2PI-effective-action_R}
\end{split}
\end{equation}
where ${S}[\phR]$ is the same as in equation~\cref{eq:classical-effective-action} with $\varphi \rightarrow \phR$, $\mu^2 \rightarrow \mu^2_{\mathrm{R}\idx{2}}$ and $\lambda \rightarrow \lRidx{4}$, and $\G_{0 \rm R}^{-1}$ is the same as~\cref{eq:free-inverse-propagator} with  $\varphi \rightarrow \phR$, $\mu^2 \rightarrow \mu^2_{\mathrm{R}\idx{0}}$ and $\lambda \rightarrow \lRidx{2}$. Moreover we defined the classical counterterm action
\begin{equation}
\delta \mathcal{S}[\varphi_{{\rm R}b}] \equiv \int {\rm d}^4x \Bigg[ \frac{\delta^\idx{2}_\varphi}{2}(\partial_\mu\varphi_{{\rm R}b})^2 - \frac{1}{2}\delta_\mu^\idx{2}\varphi_{{\rm R}b}^2 - \frac{1}{4!}\delta_\lambda^\idx{4} \varphi_{{\rm R}b}^4 \Bigg]
\label{eq:classical-effective-action_dR}
\end{equation}
and the inverse classical counterterm propagator
\begin{equation}
\mathrm{i}\delta\G_{0,ab}^{-1}(x,y;\phR) \equiv - \bigg( \delta_\varphi^\idx{0}\Box_x + \delta_\mu^\idx{0} + \frac{1}{2}\delta_\lambda^\idx{2} \varphi_{{\rm R}a}^2\bigg)\delta^{(4)}(x-y)\delta_{ab},
\label{eq:free-inverse-propagator_dR}
\end{equation}
where $\delta_\varphi^\idx{I} \equiv Z_\idx{I} - 1 $ and the other effective counterterms are defined as:
\begin{subequations}
\label{eq:effcts}
\begin{align}
\delta_\lambda^\idx{0} &\equiv Z^2_\idx{0} \big(\lRidx{0}+ \delta \lambda^\idx{0}\big) - \lRidx{0},\\
\delta_\lambda^\idx{2} &\equiv Z_\idx{0}Z_\idx{2} \big(\lRidx{2}+ \delta \lambda^\idx{2}\big) - \lRidx{2},\\
\delta_\lambda^\idx{4} &\equiv Z^2_\idx{2} \big(\lRidx{4}+ \delta \lambda^\idx{4}\big) - \lRidx{4},\\
\delta_\mu^\idx{I} &\equiv Z_\idx{I} \bigl(-\mu^2_{{\rm R}\idx{I}} + \delta \mu^2_\idx{I}\bigr) + \mu^2_{{\rm R}\idx{I}}.
\end{align}
\end{subequations}
Also in the interaction term in~\eqref{2PI-effective-action_R} the renormalized couplings in the combination $\lRidx{I} + \delta^\idx{I}_{\lambda}$ follow the power of the classical field in the interaction term~\cref{eq:shifted-action}, rewritten in terms of the renormalized quantities.  

The renormalized equations of motion can now be derived from the renormalized effective action, or more directly from~\eqref{eq:eom-for-1pt-fun} and~\eqref{eq:eom-for-2pt-fun}, by writing the the non-renormalized quantities in terms of the renormalized ones:
\begin{subequations}
\begin{align}
\biggl[ Z_\idx{2}\Box_x - \mu^2_{{\rm R}\idx{2}} + \delta_\mu^\idx{2} + \frac{1}{6}\Bigl(\lRidx{4}+\delta_\lambda^\idx{4}\Bigr)\varphi_{\rm R}^2 + \frac{1}{2} \Bigl(\lRidx{2}+\delta_\lambda^\idx{2}\Bigr)\GR\bigg] \phR &= 0,
\label{eq:eom-for-one-point-function-hartree-R}
\\
\hspace{-.4em}\biggl[ Z_\idx{0}\Box_x - \mu^2_{{\rm R}\idx{0}} + \delta_\mu^\idx{0} + \frac{1}{2}\Bigl(\lRidx{2}+\delta_\lambda^\idx{2}\Bigr) \varphi_{\rm R}^2 + \frac{1}{2} \Bigl(\lRidx{0}+\delta_\lambda^\idx{0}\Bigr)\GR \bigg]\mathrm{i}\G_{\rm R}^{ab}(x,y) &= a\delta^{ab}\delta^{(4)}.
\label{eq:eom-for-two-point-function-hartree-R}
\end{align}
\end{subequations}
Here we suppressed the arguments in the local functions $\phR(x)$ and $\GR(x,x)$, as well as in $\delta^{(4)}(x-y)$, for brevity. We now proceed to determine the various counterterms appearing in these equations and in the end find the renormalized equations of motion that include the effects of quantum corrections.

\paragraph{Auxiliary renormalization conditions.}

Because the operator acting on $\Delta^{ab}_{\rm R}$ in~\cref{eq:eom-for-two-point-function-hartree-R} is independent of branch indices, we can concentrate on the time ordered component $\Delta_{\mathrm{R}}^{11}$ of the two-point function. We choose the mass-shell renormalization condition in the vacuum configuration $\varphi_{\mathrm{R}} = v_{\mathrm{R}}$, which simultaneously minimizes the effective action. That is, we set
\begin{equation}
\mathrm{i}\bigl(\G^{11}_{\rm R}\bigr)^{-1} = p^2 - m_{\mathrm{R}}^2, \quad  \frac{\rm d}{{\rm d}p^2}\mathrm{i}\bigl(\G^{11}_{\rm R}\bigr)^{-1} = 1,
\quad {\rm and} \quad \frac{\delta\Gamma_{\rm 2PI}}{\delta \varphi_{\mathrm{R}}}\Big|_{\varphi_{\mathrm{R}}=v_{\mathrm{R}}} = 0.
\label{eq:intermediate-ren-conditions}
\end{equation}
These conditions imply that $Z_\idx{0} = 1$ in the Hartree approximation, and in our current scheme we can also set $Z_\idx{2}=1$ (see footnote~\cref{fnote:physical-parameters} below). The renormalization conditions~\cref{eq:intermediate-ren-conditions} then become:
\begin{subequations}
\begin{align}
- \mu^2_{{\rm R}\idx{2}} + \delta_\mu^\idx{2} + \frac{1}{6}\Bigl(\lRidx{4}+\delta_\lambda^\idx{4}\Bigr)v_{\rm R}^2 + \frac{1}{2} \Bigl(\lRidx{2}+\delta_\lambda^\idx{2}\Bigr)\GR(v_{\rm R}) &= 0,
\label{eq:eom-phi-R}
\\
- \mu^2_{{\rm R}\idx{0}} + \delta_\mu^\idx{0} + \frac{1}{2}\Bigl(\lRidx{2}+\delta_\lambda^\idx{2}\Bigr) v_{\rm R}^2 + \frac{1}{2} \Bigl(\lRidx{0}+\delta_\lambda^\idx{0}\Bigr)\GR(v_{\rm R}\bigr) &= m_{\mathrm{R}}^2,
\label{eq:eom-delta-R}
\end{align}
\end{subequations}
where $\GR(v_{\rm R})$ refers to the still divergent local correlator computed at the renormalization point. It should be noted that $\G^{ab}_{\rm R}$ is an auxiliary function and the parameter $m^2_{\mathrm{R}}$ is not yet related to any physical mass. Similarly, none of the couplings are yet related to observables, and there is considerable amount of choice related to their definition. We will choose the following conditions:%
\footnote{These choices are partly specific for the Hartree approximation, where the self-energy $\Pi^{ab}$ is proportional to the local correlation function. In any higher order 2PI truncation $\lRidx{0}$ and $\lRidx{2}$ would need to be renormalized separately.}
\begin{subequations}
\label{eq:rencons}
\begin{align}
\delta_\lambda^\idx{0} &= \delta_\lambda^\idx{2},
\label{eq:rencons-1} \\[.5em]
-\mu^2_{{\rm R}\idx{0}} + \delta_\mu^\idx{0} &= -\mu^2_{{\rm R}\idx{2}} + \delta_\mu^\idx{2}, 
\label{eq:rencons-2} \\
\lRidx{4} &= \lRidx{2} - \frac{1}{3}\delta_\lambda^\idx{4} + \delta_\lambda^\idx{2}.
\label{eq:rencons-3}
\end{align}
\end{subequations}
Because $Z_\idx{0,2}=1$ here, equation~\cref{eq:rencons}, together with~\cref{eq:ren-cond,eq:effcts} implies that $\lRidx{0} = \lRidx{2}$. Equation~\cref{eq:rencons-2} is less restrictive: it merely states that both renormalized mass terms are related to the same bare mass term. Conditions~\cref{eq:rencons-2,eq:rencons-3} still allow us to choose $\delta_\mu^\idx{2}$ and $\delta_\lambda^\idx{4}$ such that $m_{\rm R}^2$ and $\lRidx{4}$ can be matched to a physical mass parameter and a physical coupling. Using the conditions~\cref{eq:rencons} and equation~\cref{eq:eom-delta-R} we can write equation~\cref{eq:eom-phi-R} simply as
\begin{equation}
m_{\rm R}^2 - \frac{1}{3}\lRidx{4}v_{\rm R}^2 = 0.
\end{equation}
That is, we are able to keep the tree-level relation between the coupling $\lRidx{4}$, the background field $v_{\rm R}$ and the mass parameter $m_{\rm R}^2$.

\paragraph{Cancelling the sub-divergences.}

In order to proceed, we need to find out the divergence structure of the local correlation function. Using dimensional regularization we can write
\begin{equation}
\GR(v_{\rm R}) = Q^\epsilon \int \frac{{\rm d}^dp}{(2\uppi )^d}\,\G_{\rm R}^{11} (p) =  -\frac{m_{\mathrm{R}}^2}{16\uppi^2}\biggl[\frac{2}{\overline{\epsilon}} + 1 - \ln\biggl(\frac{m^2_{\mathrm{R}}}{Q^2}\biggr)\biggr],
\end{equation}
where $\epsilon \equiv 4-d$ and $2/{\overline\epsilon} \equiv 2/\epsilon - \gamma_{\mathrm{E}} + \ln(4\uppi)$ and $Q$ is an arbitrary renormalization scale. We now separate $\GR$ into divergent and finite parts as follows:
\begin{equation}
\GR(v_{\rm R}) \equiv m_{\mathrm{R}}^2 \G_{\overline\epsilon} + \G_{\rm F0}\bigl(m_{\rm R}^2,Q\bigr),
\label{eq:division-R}
\end{equation}
where $\G_{\overline\epsilon} \equiv -1/\bigl(8\uppi^2\overline\epsilon\bigr)$. In what follows we will suppress the $Q$-dependence of the function $\Delta_{\rm F0}$ for brevity. Next we insert the decomposition~\cref{eq:division-R} back into equation~\cref{eq:eom-delta-R}, use relations~\cref{eq:rencons} and let the leading order terms define the renormalized mass independently from the terms containing divergences or counterterms. In this way we get two equations out of the equation~\cref{eq:eom-delta-R}: 
\begin{align}
m_{\rm R}^2 &\equiv -\mu_{{\rm R}\idx{2}}^2 + \frac{1}{2}\lRidx{2}
\Bigl[ v_{\rm R}^2 + \G_{\rm F0}\bigl(m_{\rm R}^2\bigr)\Bigr],
\label{eq:tree-level-def}
\\
0 &=\delta_\mu^\idx{2} + \frac{1}{2}\delta_\lambda^\idx{2}\Bigl[ v_{\rm R}^2 + \G_{\rm F0}\bigr(m_{\rm R}^2\bigr)\Bigr] + \frac{1}{2} \Bigl(\lRidx{2}+\delta_\lambda^\idx{2}\Bigr)m_{\rm R}^2\G_{\overline\epsilon}. 
\label{eq:loop-level-eq}
\end{align}
Using definition~\cref{eq:tree-level-def} again in equation~\cref{eq:loop-level-eq} and rearranging we get
\begin{equation}
\phantom{H}
\delta_\mu^\idx{2} - \frac{1}{2} \Bigl(\lRidx{2}+\delta_\lambda^\idx{2}\Bigr)\mu_{{\rm R}\idx{2}}^2\G_{\overline\epsilon}
+ \frac{1}{2}\Bigl[ v_{\rm R}^2 + \G_{\rm F0}\bigl(m_{\rm R}^2\bigr)\Bigr] \biggl[ \delta_\lambda^\idx{2} + \frac{1}{2} \Bigl(\lRidx{2}+\delta_\lambda^\idx{2}\Bigr)\lRidx{2}\G_{\overline\epsilon}\bigg] = 0. 
\label{eq:loop-level-rearr}
\end{equation}
This equation has a consistent solution where the leading constant terms and the terms multiplying the combination (the sub-divergence part) $v_{\rm R}^2+\Delta_{\rm F0}$ cancel independently. This gives two constraint equations,
\begin{subequations}
\label{eq:last-ren-eqs}
\begin{align}
\delta_\lambda^\idx{2} + \frac{1}{2} \Bigl(\lRidx{2}+\delta_\lambda^\idx{2}\Bigr)\lRidx{2}\G_{\overline\epsilon} &= 0,
\label{eq:last-ren-eqs-1} \\
\delta_\mu^\idx{2} - \frac{1}{2} \Bigl(\lRidx{2}+\delta_\lambda^\idx{2}\Bigr)\mu_{{\rm R}\idx{2}}^2\G_{\overline\epsilon} &= 0,
\label{eq:last-ren-eqs-2}
\end{align}
\end{subequations}
from which we can finally solve the counterterms $\delta_\lambda^\idx{2}$ and $\delta_\mu^\idx{2}$:
\begin{equation}
\delta_\lambda^\idx{2} = - \frac{\sfrac{1}{2}\bigl(\lRidx{2}\bigr)^2\G_{\overline\epsilon}}{1+\sfrac{1}{2}\lRidx{2}\G_{\overline\epsilon}}  
\qquad {\rm and} \qquad 
\delta_\mu^\idx{2} = \mu^2_{{\rm R}\idx{2}}\frac{\sfrac{1}{2}\lRidx{2}\G_{\overline\epsilon}}{1+\sfrac{1}{2}\lRidx{2}\G_{\overline\epsilon}}.
\label{eq:auxiliary-cts}
\end{equation}

\paragraph{Scale dependence.}

The scale dependence 
of the auxiliary couplings and the mass parameters can now be worked out as usual by requiring that the bare parameters do not run: 
$\partial_Q\bigl[Q^\epsilon\bigl(\lRidx{2} + \delta_\lambda^\idx{2}\bigr)\bigr] = 0$ and $\partial_Q\bigl[Q^\epsilon\bigl( \mu^2_{{\rm R}\idx{I}} - \delta_\mu^\idx{I}\bigr)\bigr] = 0$. Using equations~\cref{eq:auxiliary-cts} one then immediately finds:
\begin{equation}
Q\frac{\partial\lRidx{2}}{\partial Q}= \frac{\bigl(\lRidx{2}\bigr)^2}{16\uppi^2}
\qquad {\rm and} \qquad
Q\frac{\partial\mu^2_{{\rm R}\idx{2}}}{\partial Q} = \frac{\lRidx{2}\mu^2_{{\rm R}\idx{2}}}{16\uppi^2}.
\label{eq:running-equations}
\end{equation}
The latter equation applies for both mass parameters, assuming that $\delta_\mu^\idx{0}$ and 
$\delta_\mu^\idx{2}$ differ by at most a finite and \(Q\)-independent term, which is the case in the Hartree approximation. Equations~\cref{eq:running-equations} can be easily integrated:
\begin{equation}
\lRidx{2}(Q) = \frac{\lambda^\idx{2}_{{\rm R}}(Q_0)}{1+\frac{\lambda^\idx{2}_{{\rm R}}(Q_0)}{32\uppi^2}\ln\Bigl(\frac{Q^2_0}{Q^2}\Bigr)}  
\qquad {\rm and} \qquad 
\mu^2_{{\rm R}\idx{I}}(Q) = \frac{\mu^2_{{\rm R}\idx{I}}(Q_0)}
                                      {1+\frac{\lambda^\idx{2}_{{\rm R}}(Q_0)}{32\uppi^2}\ln\Bigl(\frac{Q^2_0}{Q^2}\Bigr)}.
\label{eq:running-parameters}
\end{equation}
Remember that as a result of equation~\cref{eq:rencons-1} $\lRidx{0}=\lRidx{2}$. On the other hand, the coupling $\lRidx{4}$ does not run at all; indeed, to keep $\lRidx{4}$ finite, we must have $\delta_\lambda^\idx{4} = 3\delta_\lambda^\idx{2}$ up to finite terms according to equation~\cref{eq:rencons-3}, which implies 
\begin{equation}
Q\frac{\partial\lRidx{4}}{\partial Q}= 0
\qquad \Rightarrow \qquad \lRidx{4} = \rm constant.
\label{eq:running-equations-four}
\end{equation}
We shall see below that $\lRidx{4}$ can be further fixed by some physical condition.

%
\subsection{Renormalized equations of motion}
\label{sec:renormalized-eom}
%

It is essential to impose a correct treatment of the local correlation function away from the renormalization point in the equations of motion~\cref{eq:eom-for-one-point-function-hartree-R,eq:eom-for-two-point-function-hartree-R}. Analogously to~\cref{eq:tree-level-def}, we first define a leading order mass function that contains all finite terms in equation~\cref{eq:eom-for-two-point-function-hartree-R}:
\begin{equation}
m^2(\phR,\G_{\rm F}) \equiv - \mu^2_{{\rm R}\idx{2}} + \frac{1}{2}\lRidx{2}\Bigl(\varphi_{\rm R}^2 + \G_{\rm F}\Bigr).
\label{eq:effective-mass-full}
\end{equation}
Here $\G_{\rm F}$ is the finite part of the local correlation function, which must be defined analogously to equation~\cref{eq:division-R}:
\begin{equation}
\GR \equiv m^2(\phR,\G_{\rm F}) \G_{\overline\epsilon} + \G_{\rm F}.
\label{eq:division-full}
\end{equation}
Note that both the finite part and the divergence contain non-trivial contributions from both the vacuum and the non-equilibrium fluctuations. Using definitions~\cref{eq:effective-mass-full} and~\cref{eq:division-full} we can write equation~\cref{eq:eom-for-two-point-function-hartree-R} as follows:
\begin{equation}
\begin{split}
\phantom{H}
\biggl[\Box_x + m^2(\phR,\G_{\rm F}) &+ \delta_\mu^\idx{2} + \frac{1}{2}\delta_\lambda^\idx{2}\Bigl( \varphi_{\rm R}^2 + \G_{\rm F}\Bigr) 
 \\
&+ \frac{1}{2} \Bigl(\lRidx{2}+\delta_\lambda^\idx{2}\Bigr)m^2(\phR,\G_{\rm F})\G_{\overline\epsilon}\biggr]\mathrm{i}\G^{ab}_{\rm R}(x,y) = a\delta^{ab}\delta^{(4)}(x-y).
\label{eq:loop-level-eq-full}
\end{split}
\end{equation}
Using definition~\cref{eq:effective-mass-full} again one can show that all divergent terms in equation~\cref{eq:loop-level-eq-full} arrange as in equation~\cref{eq:loop-level-rearr} and cancel as a result of the renormalization conditions~\cref{eq:last-ren-eqs}. Then, noting that $\lRidx{4}+\delta_\lambda^\idx{4} = -2\lRidx{4} + {\mathcal O}(\epsilon )$,
the same manipulations can be done also in equation~\cref{eq:eom-for-one-point-function-hartree-R}. This results in the final renormalized equations of motion:
\begin{subequations}
\begin{align}
\Big[ \Box_x + m^2(\phR,\G_{\rm F}) \Big] \phR &= \frac{1}{3}\lRidx{4}\varphi_{\rm R}^3,
\label{eq:eom-for-phi-R}
\\
\phantom{H}
\Big[ \Box_x + m^2(\phR,\G_{\rm F}) \Big]\mathrm{i}\G_{\rm R}^{ab}(x,y) &= a\delta^{ab}\delta^{(4)}(x-y).
\label{eq:eom-for-delta-R}
\end{align}
\end{subequations}
These equations appear deceivingly simple: when written for the Wightman function $\Delta^\lt_{\rm R} = \Delta^{+-}_{\rm R}$, equation~\cref{eq:eom-for-delta-R} takes the form of a wave equation with a time-dependent mass and, as we shall see in the next section, equation~\cref{eq:eom-for-phi-R} describes the motion of the one-point function in a quantum corrected effective potential including backreaction from non-equilibrium modes. However, despite their apparent simplicity, the equations are strongly coupled through the local correlator in the gap equation~\cref{eq:effective-mass-full} for the mass term.

%
\subsection{Effective potential and physical parameters}
%

Let us now consider the {\em adiabatic limit} of the evolution equations, where $\Delta_{\rm F}$ is given purely by vacuum fluctuations with no physical excitations. In this case definition~\cref{eq:effective-mass-full} reduces to
\begin{equation}
\overbar m^2(\phR) \equiv - \mu^2_{{\rm R}\idx{2}} 
+ \frac{1}{2}\lRidx{2}\Bigl[\varphi_{\rm R}^2 + \G_{\rm F0}\bigl(\overbar m^2(\phR)\bigr)\Bigr],
\label{eq:effective-mass-vacuum}
\end{equation}
and correspondingly
\begin{equation}
{\overbar \G}_{\rm R}(\phR) \equiv \overbar m^2(\phR) \G_{\overline\epsilon} + \G_{{\rm F0}}\bigl(\overbar m^2(\phR)\bigr).
\label{eq:division-bg}
\end{equation}
Note that $\overbar m^2(\phR)$ and $\overbar\Delta_{\mathrm{R}}$ differ from definitions~\cref{eq:effective-mass-full} and~\cref{eq:division-full} only through a different value of the background field $\phR$. Using the equation of motion~\eqref{eq:eom-for-two-point-function-hartree-R} in the renormalized 2PI action~\eqref{2PI-effective-action_R} we can write down the 1PI effective potential in the Hartree approximation as follows: 
\begin{equation}
V_{\rm H}(\phR) = -\frac{1}{V}\, \Gamma_{\rm 2PI}^{\rm H} \bigl(\phR, {\overbar \G}_{\mathrm{R}}\bigr) = V_0(\phR) + \frac{\mathrm{i}}{2V} \, {\rm Tr}\Bigl[\ln\bigl({\overbar\G}_{\rm R}\bigr)\Bigr] -\frac{1}{8}\Bigl(\lRidx{2}+\delta_\lambda^\idx{2}\Bigr){\overbar\G}^2_{\rm R}(\phR),
\label{2PI-effective-potential}
\end{equation}
where $V$ is the space-time volume and the tree-level effective potential is
\begin{equation}
V_0(\phR) = \frac{1}{2}\Bigl(-\mu^2_{{\rm R}\idx{2}} + \delta_\mu^\idx{2}\Bigr)\varphi_{\rm R}^2
+\frac{1}{4!}\Bigl(\lRidx{4} + \delta_\lambda^\idx{4}\Bigr)\varphi_{\rm R}^4 
= -\frac{\lRidx{4}}{12}\varphi_{\rm R}^4,
\end{equation}
where in the last step we dropped all terms of order $\epsilon$. Writing \(\mathrm{i}{\rm Tr}\Bigl[\ln\bigl({\overbar\G}_{\rm R}\bigr)\Bigr] = V \int \mathrm{d}\overbar{m}^2 \,{\overbar\G}_{\rm R}\) and using equation~\cref{eq:division-bg} one finds that the divergences cancel between the two last terms in equation~\cref{2PI-effective-potential} and the finite part of ${\rm Tr}\Bigl[\ln\bigl( {\overbar\G}_{\rm R}\bigr)\Bigr]$ creates the one-loop correction to the effective potential. After a little algebra one finds the result:
\begin{equation}
V_{\rm H}(\phR) = -\frac{\lRidx{4}}{12}\varphi^4_{\rm R} + \frac{\overbar m^4(\phR)}{2\lRidx{2}} -\frac{\overbar m^4(\phR)}{64\uppi^2}\biggl[\ln\biggl(\frac{\overbar m^2(\phR)}{Q^2}\biggr) - \frac{1}{2} \biggr].
\label{2PI-effective-potential-final}
\end{equation}
This is the vacuum effective potential in the Hartree approximation, found for example in reference~\cite{AmelinoCamelia:1992nc}. Despite the apparent $Q$-dependence, $V_{\mathrm{H}}(\phR)$ is in fact scale-independent. Indeed, one can first show from definition~\cref{eq:effective-mass-vacuum}, using~\cref{eq:running-equations}, that  $\partial_Q\overbar m^2(\phR)=0$. Then by a direct differentiation and using equations~\cref{eq:running-equations} and~\cref{eq:running-equations-four} one finds that $\partial_Q V_{\mathrm{H}}(\phR)=0$.

\paragraph{Physical parameters.}

Differentiating~\cref{eq:effective-mass-vacuum} with respect to~$\phR$ one can first derive the identity
\begin{equation}
\frac{\partial \overbar m^2}{\partial \phR}\biggl[1-\frac{\lRidx{2}}{32\uppi^2}\ln\biggl(\frac{\overbar m^2}{Q^2}\biggr)\biggr] = \lRidx{2}\phR. 
\label{eq:vacuum-mass-eq}
\end{equation}
Using~\cref{eq:vacuum-mass-eq} it is simple to show that the first derivative of the potential can be written as
\begin{equation}
\frac{\partial V_{\rm H}}{\partial \phR} = -\frac{1}{3}\lRidx{4}\varphi_{\rm R}^3 + \overbar m^2(\phR)\phR.
\label{eq:vacuum-split}
\end{equation}
Comparing equation~\cref{eq:vacuum-split} with equation~\cref{eq:eom-for-phi-R} we can see that in the case of pure vacuum fluctuations the equation of motion can be written as $\partial_t^2 \phR + \partial V_{\rm H}/\partial\phR = 0$.  Differentiating equation~\cref{eq:vacuum-split} once more with respect to $\phR$ one finds
\begin{equation}
\frac{\partial^2 V_{\rm H}}{\partial \varphi_{\rm R}^2} = \overbar m^2(\phR) + \Bigl[ \lRidx{2}\bigl(\overbar m^2(\phR)\bigr) - \lRidx{4} \Bigr]\varphi_{\rm R}^2. 
\label{eq:second-derivative-of-potential}
\end{equation}
Because $\overbar m^2(v_{\rm R}) = m_{\mathrm{R}}^2$, we now see that the on-shell mass parameter $m_{\mathrm{R}}$ of the auxiliary propagator can be identified with a physical mass,%
\footnote{\label{fnote:physical-parameters} In fact these relations imply that $m_{\mathrm{R}}$ corresponds to a mass defined at $p^2=0$, but in the Hartree case this is the same as the physical pole mass. Going beyond Hartree approximation, one can still make $m_{\rm R}$ agree with the physical on-shell mass using the remaining freedom in definitions~\cref{eq:rencons} and in the definition of the wave-function counterterm $Z_\idx{2}$, which allow adding finite parts to $\delta_\varphi^\idx{2}$, $\delta_\mu^\idx{2}$ and $\delta_\lambda^\idx{4}$.}
if we at the same time define
\begin{equation}
\lRidx{2}(m_{\mathrm{R}}) \equiv \lRidx{4}.
\label{eq:fix-lambda}
\end{equation}
This is the choice of parameters we shall use in the rest of this paper. 

So far we have defined the counterterm $\delta_\lambda^\idx{4}$ only up to a finite constant. This, and other remaining freedom in choosing the counterterms (see footnote~\cref{fnote:physical-parameters}) would allow us to further match $\lRidx{4}$ to some observable. Given that $\lRidx{4}$ does not run, equations~\cref{eq:fix-lambda,eq:second-derivative-of-potential} are enough to fix the parameters of the theory. Going beyond the Hartree approximation would lead to more complicated calculations and relations between the auxiliary parameters, but would not change the derivation conceptually.

%
\subsection{Finite temperature effective potential}
\label{sec:effpot}
%

In our derivation in section~\cref{sec:renormalized-eom} we did not specify the finite part of the local correlation function $\Delta_{\rm F}$, and in what follows we will compute it numerically from the equations of motion. Before that it is useful to make one more observation concerning thermal corrections.
Indeed, we can include thermal corrections by a simple generalization of equations~\cref{eq:effective-mass-vacuum,eq:division-bg}: 
\begin{equation}
\overbar m^2(\phR,T) \equiv - \mu^2_{{\rm R}\idx{2}} 
+ \frac{1}{2}\lRidx{2}\Bigl[\varphi_{\rm R}^2 + \overbar\G_{\rm F}(\phR,T)\Bigr],
\label{eq:effective-mass-thermal}
\end{equation}
with $\overbar\G_{\rm R}(\phR,T) \equiv \overbar m^2(\phR,T) \G_{\overline\epsilon} + \overbar\G_{\rm F}(\phR,T)$ and
\begin{equation}
\overbar\G_{\rm F}(\phR,T) \equiv \G_{\rm F0}\bigl(\overbar m^2(\phR,T)\bigr) 
+ T^2{\mathcal I}\bigl(\overbar m^2(\phR,T)/T^2\bigr),
\label{eq:division-bg-T}
\end{equation}
where ${\mathcal I}\bigl(x\bigr) = 2\partial_{x}{\mathcal J}\bigl(x\bigr)$ and ${\mathcal J}\bigl(x\bigr)$ is the dimensionless bosonic one-loop thermal integral 
\begin{equation}
{\mathcal J}(x) \equiv \frac{1}{2\uppi^2}\,{\rm Re} \int_0^{\infty} {\rm d}y \, y^2\ln\Bigl(1-\mathrm{e}^{-\sqrt{y^2+x-{\mathrm i}\varepsilon}}\Bigr).
\end{equation}
Here the infinitesimal imaginary part ${\mathrm i}\varepsilon$ defines the correct branch of the logarithm for a negative $\overbar m^2$. With these definitions one can go through the analysis of the previous paragraph and show that the equation of motion of the homogeneous field is now $\partial_t^2 \phR + \partial V_{\rm H}(\phR,T)/\partial\phR = 0$, where $V_{\rm H}(\phR,T)$ is the thermally corrected, scale independent effective potential in the Hartree approximation:
\begin{equation}
V_{\rm H}(\phR,T) = V_{\rm H}\bigl(\phR\bigr)\Big|_{m\rightarrow \overbar m_T} - \frac{1}{2}\overbar m_T^2T^2{\mathcal I}\bigl(\overbar m^2_T/T^2\bigr) + T^4{\mathcal J}\bigl(\overbar m^2_T/T^2\bigr),
\label{eq:finite-T-pot}
\end{equation}
where $\overbar m^2_T \equiv \overbar m^2(\phR,T)$. Note that in the 2PI approach also the vacuum part \(V_{\rm H}(\phR)\)
of the potential is computed with the thermally corrected mass, which is the solution to equations~\cref{eq:effective-mass-thermal,eq:division-bg-T}.              
It is worth mentioning that in each special case considered above, from the vacuum renormalization~\cref{eq:division-R} to the quantum corrected effective action with~\cref{eq:division-bg-T} and without~\cref{eq:division-bg} thermal corrections, and finally to the general case~\cref{eq:division-full}, the divergence that gets removed by counterterms is different and depends on the value of the background field, the temperature and the particle distribution. This is an unavoidable feature of the 2PI equations with a finite order truncation. However, in all cases the counterterms themselves remain the same, uniquely defined constants. 

\paragraph{Comparison to ring-resummed potentials.}

Equations~\cref{eq:effective-mass-thermal,eq:division-bg-T} and~\cref{eq:finite-T-pot} provide a consistent resummation of the thermal potential to super-daisy level. They can be seen as a consistent generalization of the Parwani resummation method~\cite{Parwani:1991gq}. In these approaches the thermal corrections affect all modes on equal footing, while in the usual ring resummation method~\cite{Carrington:1991hz,Arnold:1992rz} only the long wavelength modes are screened by the short wavelength modes in a thermal plasma. The advantage of the potential~\cref{eq:finite-T-pot} is that it provides a consistently renormalized, smooth continuation between the non-relativistic and relativistic regimes. In all other ring-resummed potentials this behaviour has to be put in by hand.

To effect a fair comparison of different approximations, we write all potentials using the same renormalization conditions. To be concrete, we use the conditions $\partial_{\varphi}^2 V(v_{\rm R}) = m_{\mathrm{R}}^2$ and $\partial_{\varphi}^4 V(v_{\rm R}) = \lambda_{\rm R}$. With these conditions the standard one-loop corrected potential without the ring-corrections becomes
\begin{equation}
V_{\rm 1L}(\varphi_{\mathrm{R}},T) \equiv -\frac{1}{2}\mu_{\rm R}^2\varphi_{\mathrm{R}}^2 + \frac{1}{4!}\lambda_{\rm R}\varphi_{\mathrm{R}}^4 
+ V_{\rm 1-loop}(\varphi_{\mathrm{R}}) + T^4\mathcal{J}\biggl(\frac{m_0^2(\varphi_{\mathrm{R}})}{T^2}\biggr),
\label{eq:VCW}
\end{equation}
where $m_0^2(\varphi_{\mathrm{R}}) = -\mu_{\rm R}^2 + \frac{1}{2}\lambda_{\rm R}\varphi_{\mathrm{R}}^2$ and the standard one-loop vacuum potential is (this potential also satisfies the condition $\partial_{\varphi}V_{\rm 1-loop}(v_{\rm R}) = 0$)
\begin{equation}
V_{\rm 1-loop}(\varphi_{\mathrm{R}}) = 
\frac{1}{64\uppi^2} \biggl\{ m_0^4(\varphi_{\mathrm{R}})\biggl[ \ln\biggl(\frac{m_0^2(\varphi_{\mathrm{R}})}{m^2_{\rm R}}\biggr) - \frac{3}{2} \biggr] + 2m^2_{\rm R} m_0^2(\varphi_{\mathrm{R}}) \biggr\}.
\end{equation}
In the Parwani approximation~\cite{Parwani:1991gq} one replaces $m_0^2(\varphi_{\mathrm{R}})$ with the lowest order thermal mass
$m_0^2(\varphi_{\mathrm{R}},T)=m_0^2(\varphi_{\mathrm{R}})+\frac{1}{24}\lambda_{\rm R}T^2$ in equation~\cref{eq:VCW} and in the ring approximation~\cite{Carrington:1991hz,Arnold:1992rz}, where only the zero-mode is dressed by thermal corrections, one finds:
\begin{equation}
V_{\rm Ring}(\phR,T) \equiv V_{\rm 1L}(\phR,T) 
+ \frac{T}{12\uppi} \mathrm{Re} \Big(m_0^3(\phR) - m_0^3(\phR,T)\Big).
\end{equation}
Above we wrote the Hartree potential in terms of the scale dependent variables. However, since the potential is actually scale independent, we can rewrite it at the scale $Q=m_{\rm R}$, explicitly in terms of the physical parameters:
\begin{equation}
V_{\rm H}(\phR,T) \!=\! -\frac{\lRidx{4}}{12}\varphi_{\rm R}^4 
+ \frac{\overbar m^4_T}{2\lRidx{4}} -\frac{\overbar m^4_T}{64\uppi^2}\biggl[\ln\biggl(\frac{\overbar m^2_T}{m_{\rm R}^2}\biggr) \!-\frac{1}{2} \biggr]
- \frac{\overbar m_T^2T^2}{2}{\mathcal I}\biggl(\frac{\overbar m^2_T}{T^2}\biggr) + T^4{\mathcal J}\biggl(\frac{\overbar m^2_T}{T^2}\biggr),
\label{2PI-effective-potential-physpm}
\end{equation}
where $\overbar m_T^2$ is the solution to the gap equation, which now becomes
\begin{equation}
\overbar m^2_T \equiv m_{\rm R}^2 + \frac{1}{2} \lRidx{4}(\varphi_{\rm R}^2 - v_{\rm R}^2)+ \frac{\lRidx{4}}{32\uppi^2}
\biggl[\overbar m^2_T\ln\biggl(\frac{\overbar m^2_T}{m_{\rm R}^2} \biggr) + m_{\rm R}^2 -  \overbar m^2_T\biggr] 
+ T^2{\mathcal I}\biggl(\frac{\overbar m^2_T}{T^2}\biggr),
\label{eq:effective-mass-thermal-physpm}
\end{equation}
with $m_{\rm R}^2 = \sfrac{1}{3}\lRidx{4}v_{\rm R}^2$ and where finally $\lRidx{4}$ is related to the renormalized coupling $\lambda_{\mathrm{R}} \equiv \partial_{\varphi}^4V_{\rm H}(v_{\mathrm{R}},0)$ by
\begin{equation}
\lambda_{\mathrm{R}} = \lRidx{4}\biggl[1 + 3\biggl(\frac{3\lRidx{4}}{32\uppi^2}\biggr) + 3\biggl(\frac{3\lRidx{4}}{32\uppi^2}\biggr)^{\!\! 2}\,\biggr],
\label{eq:effective-coupling}
\end{equation}
as can be shown by direct differentiation of equation~\cref{2PI-effective-potential-physpm}. 
%
%
\begin{figure}[t]
\begin{center}
\includegraphics[width=0.342\textwidth]{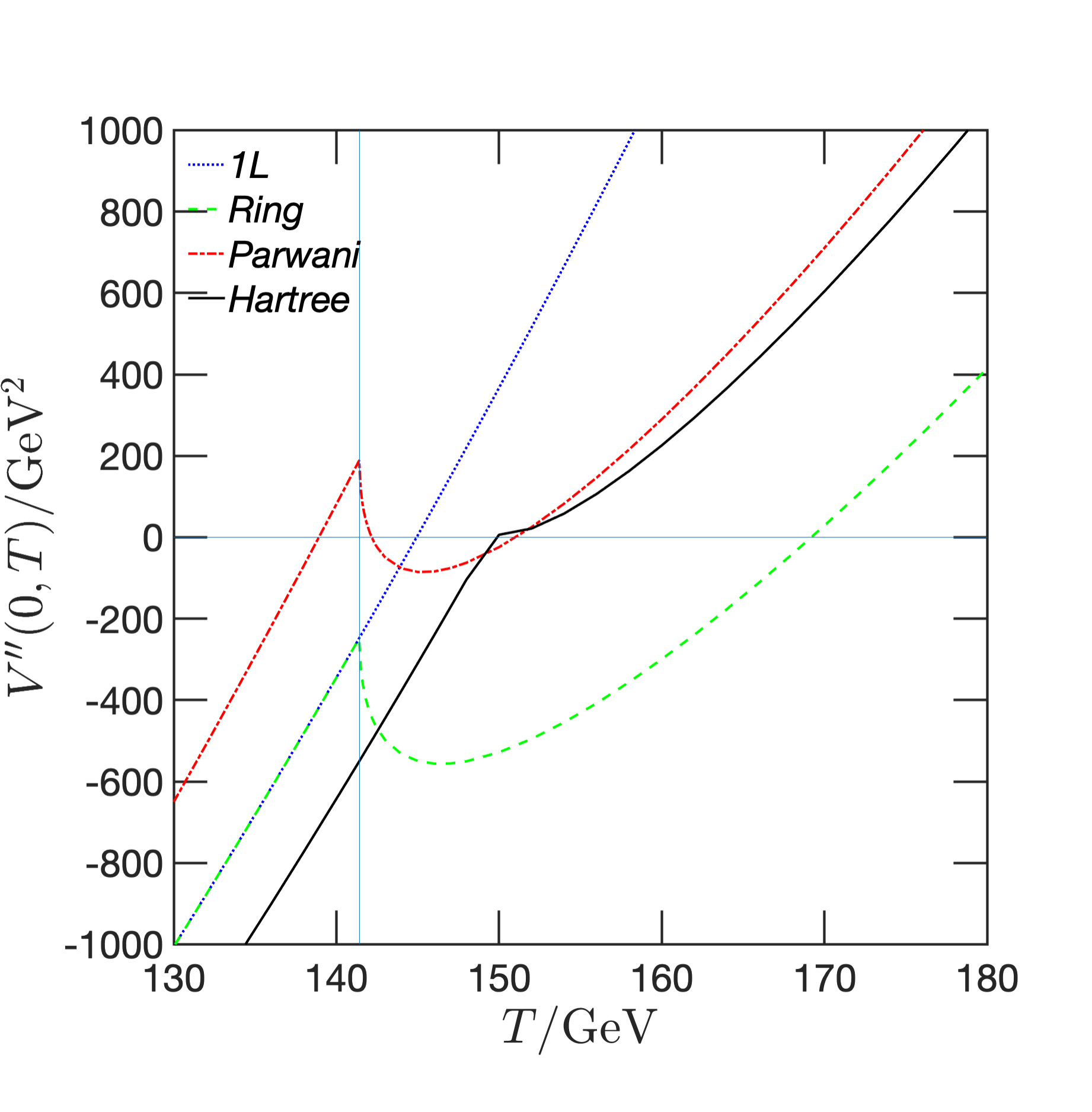}
\hskip -3mm
\includegraphics[width=0.333\textwidth]{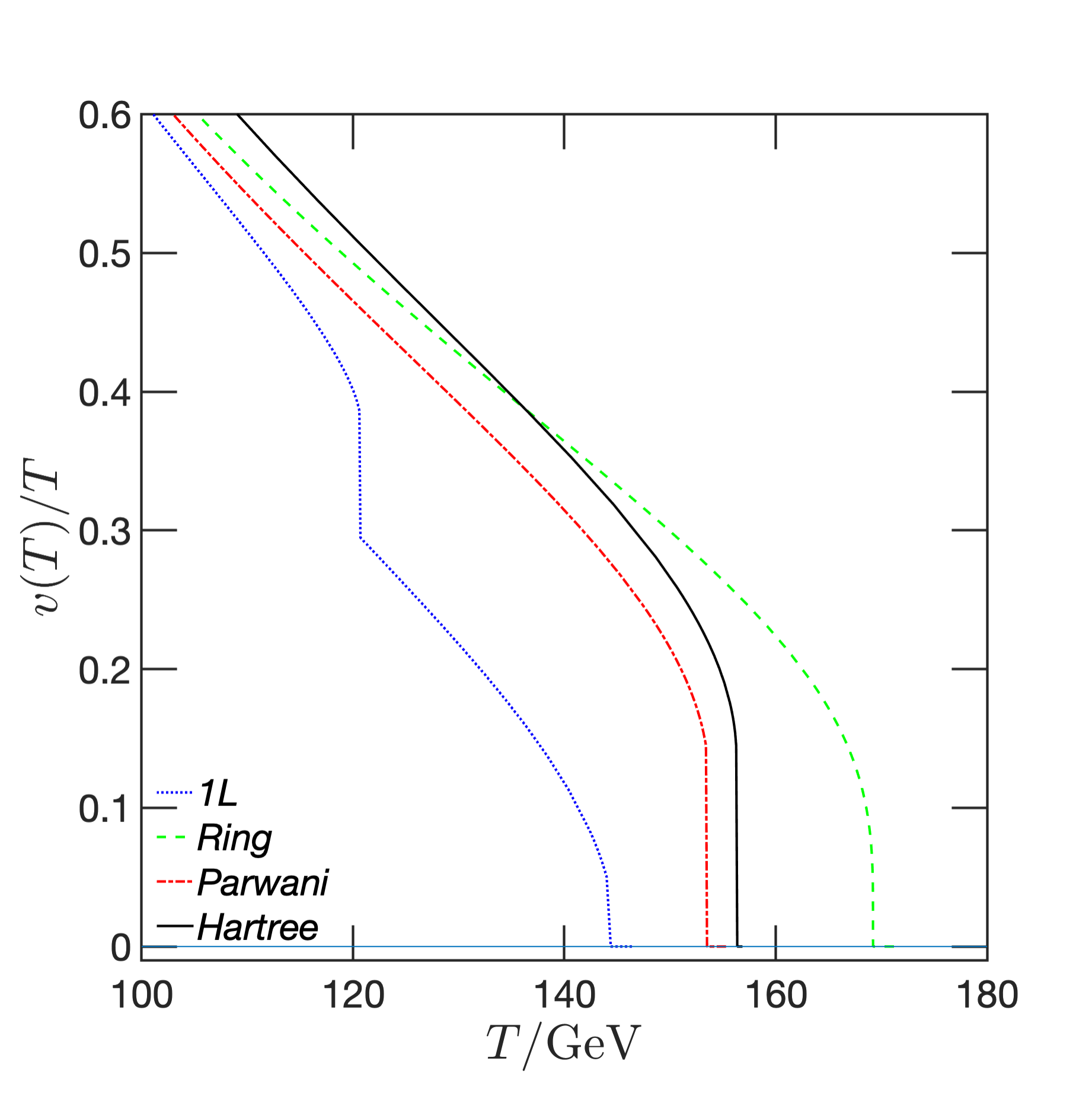}
\hskip -3mm
\includegraphics[width=0.342\textwidth]{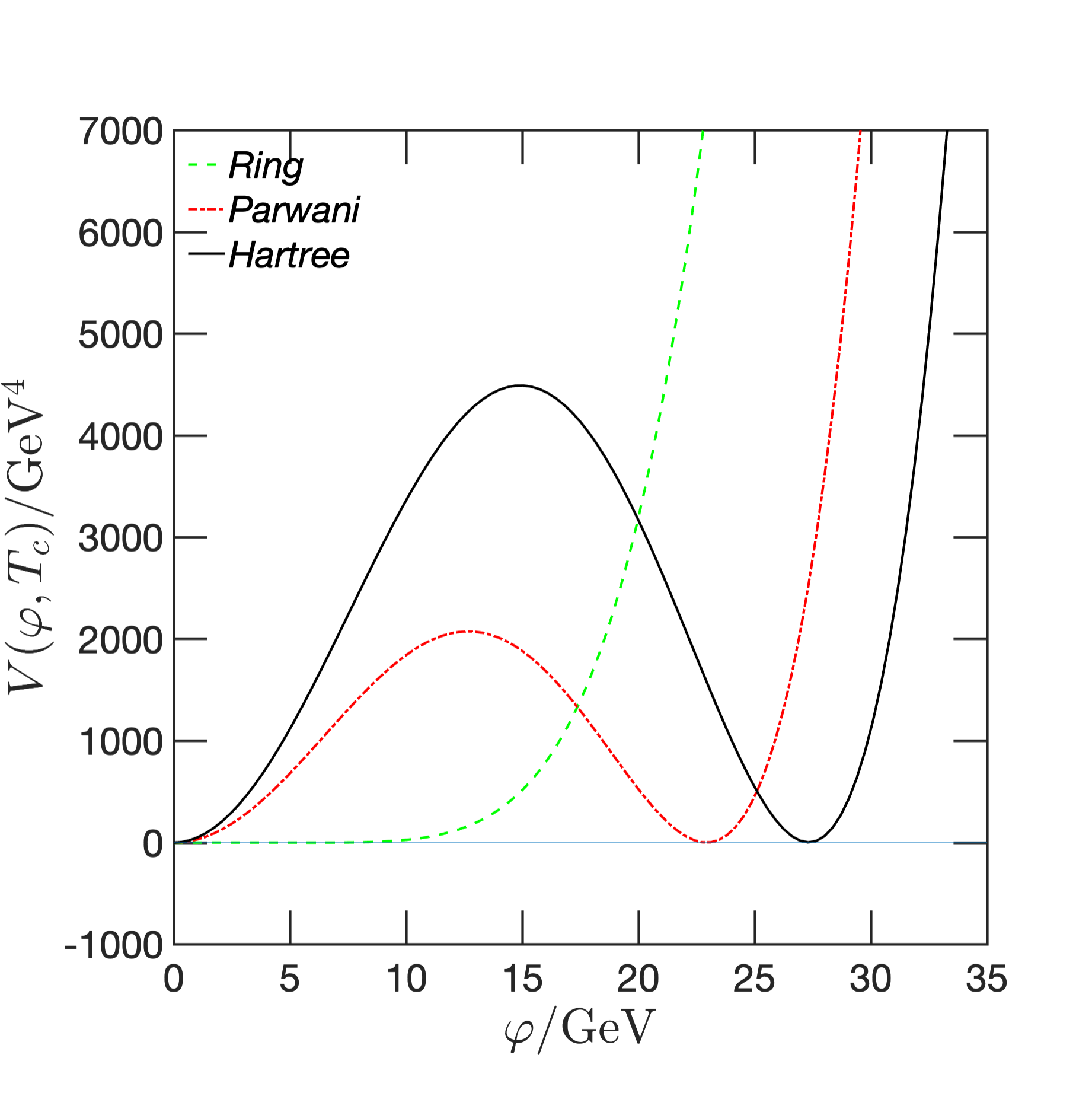}
\end{center}
\vskip-0.5cm
\caption{Left: the evolution of the second \(\varphi\)-derivatives of the different potentials at $\varphi_{\mathrm{R}}=0$. Middle: the evolution of the ratio $v(T)/T$, where $v(T)$ is the position of the second minimum. Right: the potential at critical temperature in each approximation. The critical temperatures are $T_{\mathrm{c}}\approx 169.20$ GeV in the ring, $T_{\mathrm{c}}\approx 153.29$ GeV in the Parwani and $T_{\mathrm{c}}\approx 155.67$ GeV in the Hartree approximation. We used $m_{\rm R} = 100$ GeV and $\lambda_{\rm R} = 6$ (which implies $\lRidx{4} \simeq 5.2$). The vertical line in the left panel shows $T_0 = \sqrt{12/\lambda_{\rm R}}m_{\mathrm{R}}$, where the high-temperature limit approximated thermal mass vanishes at $\varphi_{\mathrm{R}}=0$.}
\label{fig:critical_temp}
\end{figure}
%

In the left panel of figure~\cref{fig:critical_temp} we show the evolution of the second \(\varphi\)-derivatives of the potentials near the critical temperature at $\varphi_{\mathrm{R}}=0$. The sharp kinks seen in the ring (green dashed line) and Parwani (red dash-dotted line) cases at $T=T_0$ result from the non-analytic dependence of the resummed potentials on the thermally corrected mass term (we are using the high-temperature expansion for $m^2(\varphi,T)$ in these schemes). The one-loop result (blue dotted line) does not share this feature, because there we are using the non-resummed mass term. Interestingly, the Hartree result (black line) does not show signs of similar non-analyticity. In the middle panel we show the evolution of the ratio $v(T)/T$, where $v(T)$ is the position of the asymmetric minimum as a function of $T$. There are significant differences between the approximations: in all resummed potentials a metastable minimum emerges, and it has the largest jump in the Hartree case. In the one-loop case the metastability does not develop, but there is a jump in $v(T)/T$ at $T\approx 120$ GeV due to the non-analytic behaviour, now of the vacuum mass term as a function of $\varphi$. In the right panel we show the potentials at the critical temperature (whose value for each model is given in the figure caption). The transition strength is dramatically different in the different approximations and it is by far the strongest in the Hartree case. Of course one should keep in mind that this is a very simple model, with only a single scalar field. However, when one compares the one-loop results with lattice calculations, one typically finds that both ring and Parwani approximations give weaker transitions than the lattice or 3d-perturbation theory calculations~\cite{Kainulainen:2019kyp}. It would be interesting to see if the Hartree approximation was in better agreement with these schemes when applied in more complex models.

%
\section{Wigner space equations}
\label{sec:wigner}
%

We now proceed to solving the general equations~\cref{eq:eom-for-delta-R,eq:eom-for-phi-R} for homogeneous non-equilibrium systems. Of these, equation~\cref{eq:eom-for-phi-R} is already in the desired form, when we assume that field $\phR$ is homogeneous, but equation~\cref{eq:eom-for-delta-R} for the correlation function will be easier to handle in the mixed representation. Because of the homogeneity an ordinary Fourier transformation is sufficient for the spatial coordinates, but for the time variable we need the Wigner transformation:
\begin{equation}
\G_{{\rm R}\bm{k}}^{ab}(k_0,t) = \int{\rm d}r_0 \, \G_{{\rm R}\bm{k}}^{ab}\biggl(t+\frac{r_0}{2},t-\frac{r_0}{2}\biggr)\mathrm{e}^{\mathrm{i}k_0r_0},
\end{equation}
where $t \equiv \frac{1}{2}(x_0+y_0)$ and $r_0 \equiv x_0-y_0$. Because all correlation functions $\G^{ab}(x,y)$ have the same local limit, it suffices to consider the equation for the lesser Wightman function $\G^{+-}\equiv \G^\lt$. Starting from equation~\cref{eq:eom-for-delta-R}, we find that in the Wigner representation it satisfies the equation,
\begin{equation}
  \left[\sfrac{1}{4}\partial_t^2-k^2-{\mathrm i}k_0\partial_t + \mathrm{e}^{-\frac{\mathrm{i}}{2}{\partial_t^m}\partial_{k_0}}{m^2}\big(\phR, \GR\big)\right] \G_{{\rm R}\bm{k}}^\lt(k_0,t) = 0.
  \label{eq:wigner-equation-lt}
\end{equation}
Here the index $m$ in the derivative $\partial_t^m$ signals that the time-derivative acts only on the mass function and not on the propagator. Equation~\cref{eq:wigner-equation-lt} is still equivalent to~\cref{eq:eom-for-delta-R} and highly complicated because of the infinite tower of \(t\)- and \(k_0\)-derivatives involved. It can be recast into a simpler form by introducing a moment expansion. Following reference~\cite{Herranen:2008di} we first introduce the moment functions:
\begin{equation}
\rho_{n\bm{k}}(t) = \int\frac{\mathrm{d}k_0}{2\uppi}\,k_0^n\,\G^<_{\mathrm{R}\bm{k}}(k_0,t).
\label{eq:moment}
\end{equation}
Then taking the real and imaginary parts of equation~\cref{eq:wigner-equation-lt} integrated over $k_0$ and the imaginary part of the same equation integrated over $k_0$ and weighted by $k_0$, we get three equations coupling the moments $\rho_{n\bm{k}}$ with $n=0,1,2$ to the field equation for a homogeneous field $\phR(t)$:
\begin{subequations}
\label{eq:moments}
\begin{align}
 \frac{1}{4}\partial_t^2\rho_{0\bm{k}} - \rho_{2\bm{k}} + \omega^2_{\bm k}(t) \,\rho_{0\bm{k}} &= 0,
\label{eq:moments-rho0}
\\[.2em]
\partial_{t}\rho_{1\bm{k}} &= 0,
\label{eq:moments-rho1} 
\\[.2em]
\partial_{t}\rho_{2\bm{k}} -\frac{1}{2}\Bigl[\partial_t \bigl(m^2_{\rm eff}(t)\bigr) \Bigr] \rho_{0\bm{k}} &= 0,
\label{eq:moments-rho2} 
\\
\phantom{\frac{1}{2}}\Bigl[\partial_t^2 + m^2_{\rm eff}(t)\Bigr]\phR &= \frac{1}{3}\lambda^\idx{2}_{\rm R}\varphi_{\rm R}^3.
\label{eq:moments-phi}
\end{align}
\end{subequations}
We used the shorthand $m^2_{\rm eff}(t)\equiv m^2(\phR,\GR)$ for the mass defined in~\cref{eq:division-full,eq:effective-mass-full} and defined $\omega^2_{\bm{k}}(t) \equiv \bm{k}^2+m^2_{\rm eff}(t)$.  The gap equation~\cref{eq:effective-mass-full}, including the out-of-equilibrium (or thermal) modes, can be written explicitly as
\begin{equation}
m^2_{\rm eff}(t) = - \mu^2_{{\rm R}\idx{2}} 
+ \frac{1}{2}\lRidx{2}\Biggl\{\varphi_{\rm R}^2 + \Delta_{\rm F0}\bigl(m^2_{\rm eff}(t)\bigr) + 
\int_{\bm k} \nolimits \Biggl[ \rho_{0\bm{k}}(t) - \frac{\theta\bigl(\omega_{\bm k}^2(t)\bigr)}{2\omega_{{\bm k}}(t)}\Biggr] \Biggr\},
\label{eq:meff-gap-equation}
\end{equation}
where we defined \(\int_{\bm k} \equiv \frac{1}{2\uppi^2}\int_0^{\infty} {\rm d}|{\bm k}| |{\bm k}|^2\), and the Heaviside theta-function $\theta\bigl(\omega_{\bm k}^2(t)\bigr)$ ensures that the vacuum does not contain the unstable spinodal modes.%
\footnote{Spinodal modes are the unstable modes that appear when the effective mass function is negative. We define them explicitly in equation~\cref{eq:spinodal_modes} below. Note that the vacuum energy integral in the spinodal region, computed with the Heaviside function, is identical with the integral computed taking the absolute value of the mass squared function and integrating over all momenta.}

If $\rho_{0\bm{k}}(t)$ is identified with a thermal distribution (including the vacuum part), equation~\cref{eq:meff-gap-equation} clearly reduces to~\cref{eq:effective-mass-thermal}. After discretizing the momentum variable, equations~\cref{eq:moments-rho2,eq:moments-rho0,eq:moments-phi,eq:meff-gap-equation} can be written as a closed set of coupled first order differential equations, which include backreaction from the non-equilibrium modes into the evolution of the homogeneous field $\phR$. The gap equation~\cref{eq:meff-gap-equation} must be solved first at the entry to the routine, after which the solution is advanced using~\cref{eq:moments-rho2,eq:moments-rho0,eq:moments-phi}. In practice one must introduce a UV-cutoff for the magnitude of the momentum $|\bm {k}|$, but results should not depend on its precise value, because all non-trivial physics results from gradient terms acting in the infrared region. We have indeed shown that this is the case in our numerical examples.

\paragraph{Friction.}
 
Our main goal is to study the dynamical evolution of $\phR$ including the backreaction from the modes excited during the zero-crossings (parametric resonance) and from the unstable modes (spinodal, or tachyonic, instability). We would also like to study dissipative interactions in our system. To do this correctly, we should go beyond the Hartree approximation. This would be in principle a straightforward but very laborious task. Some formal results can be found for example in~\cite{PhysRevD.66.045008}. Here we will instead add phenomenological friction terms to our equations. Following references~\cite{Herranen:2008di} and~\cite{Herranen:2010mh}  we generalize equations~\cref{eq:moments-rho0,eq:moments-rho1,eq:moments-rho2} as follows:
\begin{subequations}
\label{eq:moments-fric}
\begin{align}
\frac{1}{4}\partial_t^2\rho_{0\bm{k}} - \rho_{2\bm{k}} + \omega_{\bm{k}}^2(t) \rho_{0\bm{k}} &= -{c_1}\partial_t \rho_{0\bm{k}}, 
\label{eq:moments-rho0-fric}
\\
\partial_t \rho_{1\bm{k}} &= -{c_2}\bigl(\delta\rho_{1\bm{k}} - \delta\rho_{1\bm{k}}^{\mathrm{eq}}\bigr), 
\label{eq:moments-rho1-fric}
\\
\partial_{t}\rho_{2\bm{k}} -\frac{1}{2}\Bigl[\partial_t \bigl(m^2_{\rm eff}(t)\bigr) \Bigr] \rho_{0\bm{k}} 
&= -{c_2}\bigl(\delta\rho_{2\bm{k}} - \delta\rho_{2\bm{k}}^{\mathrm{eq}}\bigr),
\label{eq:moments-rho2-fric}
\end{align}
\end{subequations}
where $\delta \rho_{n\bm{k}} \equiv \rho_{n\bm{k}} - \rho_{n\bm{k}}^{\rm vac}$ with $\rho_{n\bm{k}}^{\rm vac}$ being the vacuum moments defined in equations~\cref{eq:non-coherent-vacuum} below, and the explicit forms for the equilibrium distributions $\rho_{n{\bm k}}^{{\rm eq}}$ have to be provided externally depending on the problem. Collision integrals could be computed accurately in the context of the cQPA formalism following reference~\cite{Herranen:2010mh} (see also~\cite{Millington:2012pf}), but here we are only interested in qualitative effects, for which the above phenomenological approach is sufficient. Even then the coefficients $c_{i}$ could be some momentum dependent functions, but for simplicity we will assume that they are constants. Note that $\rho_{n\bm k}$ and $\rho_{n{\bm k}}^{{\rm eq}}$ in general have different vacuum distributions due to different respective solutions to mass gap equations. 

\paragraph{Number densities and coherence function.}

We can get a better understanding of the physical meaning of the moments by comparing them with the spectral cQPA solutions found in reference~\cite{Herranen:2008di}. As explained in section 4.2 of reference~\cite{Herranen:2008di}, the moments are in a one-to-one correspondence with the cQPA mass-shell functions $f^m_{{\bm k}\pm}$ and the coherence function $f^c_{\bm k}$. The former can be further related to the particle and antiparticle number densities $n_{\bm k}$ and $\overbar n_{\bm k}$, so that one eventually finds~\cite{Herranen:2008di,Herranen:2010mh}:
\begin{subequations}
\label{f-rho_HOM}
\begin{align}
n_{\bm{k}} &= \frac{1}{\omega_{\bm k}}\rho_{2\bm{k}} + \rho_{1\bm{k}},
\label{f-rho_HOM-n}\\
\overbar{n}_{\bm{k}} &= \frac{1}{\omega_{\bm k}}\rho_{2\bm{k}} - \rho_{1\bm{k}} - 1,
\label{f-rho_HOM-nbar}\\
f^{c\pm}_{\bm k} &= \omega_{\bm k}\rho_{0\bm{k}} - \frac{1}{\omega_{\bm k}}\rho_{2\bm{k}} 
\pm \frac{\mathrm{i}}{2}\partial_t\rho_{0\bm{k}}.
\label{f-rho_HOM-fc}
\end{align}
\end{subequations}
The coherence functions $f^{c\pm}_{\bm k}$ measure the degree of quantum coherence, or squeezing, between particle-antiparticle pairs with opposite 3-momenta~\cite{Fidler:2011yq}. A {\em non-coherent} vacuum state must then be defined as a state with no squeezing in addition to having no particles. This corresponds to setting $n_{\bm k} = \overbar n_{\bm k} = f^{c\pm}_{\bm k} \equiv 0$, which is equivalent to: 
\begin{equation}
\rho_{0\bm k}^{\rm vac} = \frac{\Theta_{\bm k}}{2\omega_{\bm k}}, \qquad 
\partial_t\rho_{0\bm k}^{\rm vac}=0, \qquad
\rho_{1\bm k}^{\rm vac} = -\frac{1}{2}
\quad \mathrm{and} \quad 
\rho_{2\bm k}^{\rm vac} = \frac{\omega_{\bm k}}{2}\Theta_{\bm k},
\label{eq:non-coherent-vacuum}
\end{equation}
where $\Theta_{\bm k} \equiv \theta\bigl(\omega_{\bm k}^2(t)\bigr)$. Because we are assuming that $\phR$ is a real scalar field we also have $\overbar n_{\bm k} = n_{\bm k}$, which implies that $\rho_{1\bm k} = -1/2$ at all times, so that the equation for $\rho_{1\bm{k}}$ is actually redundant. This is indeed a consistent solution even with the friction terms included.
%
%

%
\section{Numerical results}
\label{sec:results}
%

We shall now solve the coupled dynamical equations~\cref{eq:moments-rho0-fric,eq:moments-rho1-fric,eq:moments-rho2-fric,eq:meff-gap-equation,eq:moments-phi} in a few examples chosen to illustrate the rich physics of the strongly coupled system including the quantum backreaction. We will uncover some known results and find new phenomena associated with spinodal and resonant particle production at phase transitions%
\footnote{The use of the term phase transition is not very accurate here, as we do not have a phase transition in the same sense as for example in the electroweak transition. Rather, we have a situation where the universe evolves from a cold initial state to a hot final state. It is a common practice however to refer to this phenomenon as a phase transition as well, and we will also do so in what follows.}. 
We will show that a strong spinodal instability can cause a quantum assisted barrier penetration without tunneling, and we emphasize the difficulty of giving any sensible definition for the effective potential in a non-equilibrium system. Eventually, we will follow the full thermalization process of a scalar field starting at rest in the vacuum potential until the end, when the energy in the field is almost completely transformed into thermal fluctuations.%
\footnote{Let us make a note on units: in section~\cref{sec:effpot}, when discussing the thermal effective potentials, we gave the mass parameter a value characteristic for the electroweak phase transition, $m_{\rm R} = 100$ GeV. Below we continue to use the same value as a benchmark, and we shall be measuring all dimensionful quantities in the GeV-units. In particular, we will be measuring time in units GeV$^{-1}$, while we will be suppressing time-units in all plots. However, in all examples that we will consider below, the physical mass $m_{\rm R}$ is the only mass scale in the problem. Thus, all results are in fact valid as such for an arbitrary mass value, if only one rescales all dimensionful parameters by a suitable power of $m_{\rm R}/$GeV.}

%
\subsection{Particle production and reheating via parametric resonance}
\label{sec:parmetric-resoanance}
%

We first consider a case where the field starts from a relatively large value and oscillates several times between positive and negative field values. Because we are also interested in the spinodal instability, we consider a tree-level potential with a negative mass term. As physical parameters we use $m_{\mathrm{R}} = 100$ GeV and $\lRidx{4} = 1$, given which, $\mu_{\rm R\idx{2}}^2(Q_0)$ can be solved from~\cref{eq:tree-level-def}, while the running couplings and masses are defined in~\cref{eq:running-parameters}. We compute the initial value for the effective mass function $m^2(\phR,\GR)$ using the vacuum Hartree approximation~\cref{eq:effective-mass-vacuum}.  We used running parameters everywhere in our calculations. This served as a useful consistency check, since all final results must be (and indeed were) scale independent. In this example we also set the friction terms to zero, $c_i=0$.

\begin{figure}[t]
\begin{center}
\includegraphics[width=0.9\textwidth]{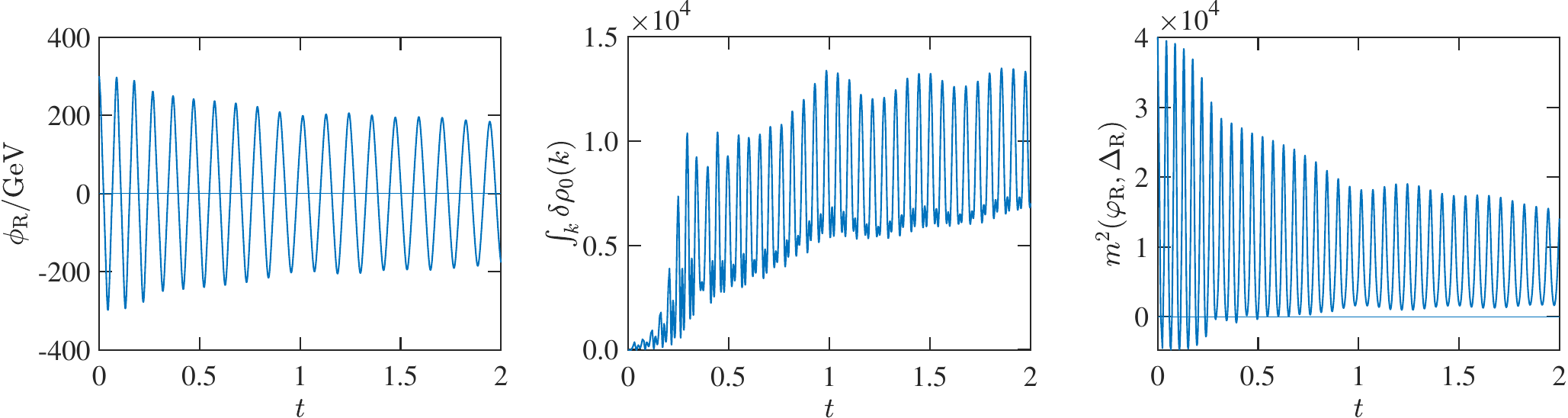}
\end{center}
\caption{Shown is the evolution of the classical field as a function of time (left), evolution of the integrated non-equilibrium part of the local correlation function (middle), and the effective mass function $m^2(\phR,\GR)$ (right). We used $\lRidx{4} = 1$, $m_{\rm R}=100$ GeV, $\varphi_{\mathrm{R,in}}=300$ GeV and $\partial_t \varphi_{\rm R,in}=0$.  The moment functions were initialized to the non-coherent vacuum values~\cref{eq:non-coherent-vacuum}. We also assumed no friction, setting $c_i$ to zero.}
\label{fig:figure4}
\end{figure}

\begin{figure}[t]
\begin{center}
\includegraphics[width=0.95\textwidth]{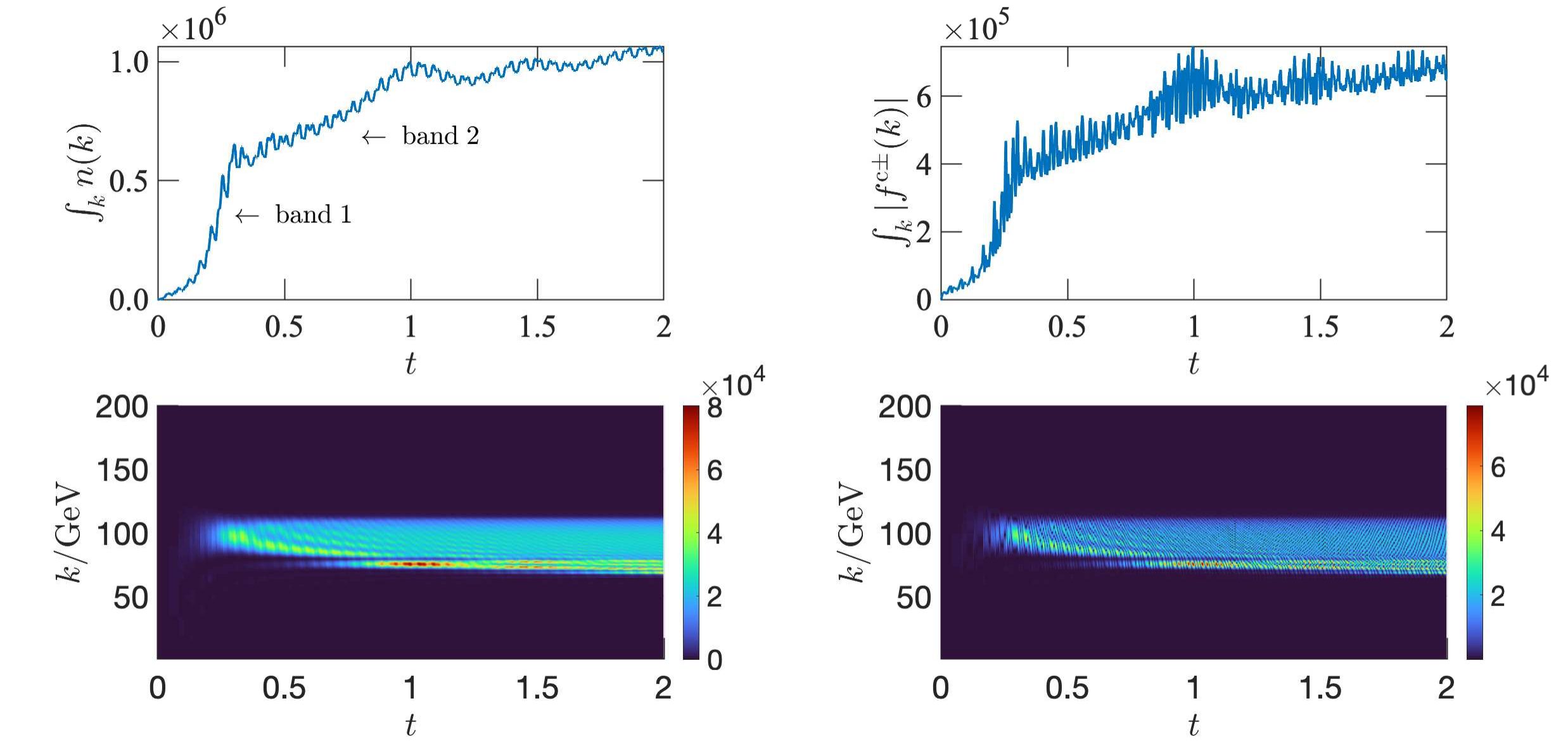}
\end{center}
\caption{Shown is the evolution of the integrated number density (top left) and the absolute value of the integrated coherence function $\bigl|f_{\bm k}^{\rm c\pm}\bigr|$ (top right), defined in equations~\cref{f-rho_HOM}, for the same parameters as in figure~\cref{fig:figure4}. The bottom row shows the heat plots in the momentum and time variables for the unintegrated distributions multiplied by the phase space factors: $\frac{{\bm k}^2}{2\uppi^2}n_{\bm k}$ (lower left) and $\frac{{\bm k}^2}{2\uppi^2}\bigl|f^{\rm c\pm}_{\bm k}\bigr|$ (lower right).}
\label{fig:figure5}
\end{figure}

The essential results for this run are shown in figures~\cref{fig:figure4,fig:figure5}. In the left panel of figure~\cref{fig:figure4} we show the evolution of the classical field $\phR$, which here displays an orderly oscillation pattern with a slowly decaying amplitude. The middle panel of figure~\cref{fig:figure4} shows the evolution of the fluctuations in the zeroth moment integrated over the 3-momentum, which is the non-equilibrium contribution to the local correlation function: $\int_{\bm k}\delta \rho_{0\bm{k}} = \int_{\bm k}\bigl(\rho_{0\bm{k}} - \rho_{0\bm{k}}^{{\rm vac}}\bigr) \equiv \delta \Delta_{\rm F}(t,t)$. These results are in good agreement with reference~\cite{PhysRevD.65.065019}, where this problem was studied earlier using the mode equation approach. The rapid increase of $\delta \Delta_{\rm F}(t,t)$ at early times is caused by two non-perturbative processes, the spinodal instability and the parametric resonance.

%
%

\paragraph{Spinodal instability.}

The presence of a spinodal instability is manifest in the right panel of figure~\cref{fig:figure4}, where the effective mass term $m^2(\phR,\GR)$ is seen to become periodically negative in the region $t\lesssim 0.25$. Indeed, whenever the mass-function is negative,  all ${\bm k}$-modes satisfying
\begin{equation}
{\bm k}^2 + m^2(\phR,\GR) < 0
\label{eq:spinodal_modes}
\end{equation}
are unstable and can grow exponentially. This is the \emph{spinodal} or \emph{tachyonic} instability. One might then be tempted to associate the growth in fluctuations in the period $t \lesssim 0.25$ fully to the spinodal instability. If this was true, the excited modes should satisfy the condition~\cref{eq:spinodal_modes}, which here translates to $|{\bm k}| \lesssim 60$ GeV. However, from figure~\cref{fig:figure5} we see that this is not the case. The fast production of modes is clearly visible in the upper panels which show the integrated particle number (left) and the integrated modulus of the coherence functions (right). But from the lower panels, showing time-momentum heat plots of the same quantities, we see that the excited modes are concentrated on a frequency band which lies entirely above the spinodal region~\cref{eq:spinodal_modes}.

\paragraph{Parametric resonance.}

While our equations are highly non-linear and strongly self-coupled, it is apparent that the structures seen in the heat plots in figure~\cref{fig:figure5} correspond to Mathieu instabilities associated with parametric resonance, familiar from the studies of inflationary reheating~\cite{kofman:1997yn}. This problem was also studied using 2PI methods in reference~\cite{PhysRevLett.91.111601}, albeit with a different set of approximations and a different potential. If we identify the mass squared of the mode function in the Mathieu equation with our mass function $m^2(\phR,\GR)$, and follow the analysis of section V in reference~\cite{kofman:1997yn}, we can (very roughly) estimate the Mathieu equation $q$-parameter in our case to be 
\begin{equation}
q \sim 2 \frac{\Delta m^2_{\rm eff}}{(2\uppi \nu)^2} \approx 2,
\label{eq:q-equation}
\end{equation}
where $\Delta m^2_{\rm eff} \approx 2\times 10^4$ GeV$^2$ is the instantaneous amplitude and $\nu \approx 21$ GeV is the local frequency of oscillations of the effective mass term $m^2(\phR,\GR)$, shown in figure~\cref{fig:figure4}. The value of the $q$-parameter, which remains roughly the same throughout the calculation, suggests an intermediate resonance between the narrow and broad regimes. Similarly, the expected position of the first resonance band is by and large estimated to be 
\begin{equation}
|\bm{k}|_{\rm rb}\sim \frac{\uppi \nu}{\sqrt[4]{2}} \approx 60 \,\mathrm{GeV}.
\end{equation}
This result, and the expected width of the resonance~\cite{kofman:1997yn} $\smash\Delta |\bm{k}| \sim |\bm{k}|_{\rm rb} \approx 60$ GeV are in qualitative agreement with our results. In  figure~\cref{fig:figure5} we can even observe a second, much narrower band below the first one, which dominates the particle production at $t \approx 1$. While this is again in agreement with the qualitative expectations, its interpretation via Mathieu equation methods becomes even more tenuous. At late times $t \gtrsim 0.3$ the shape of the growth pattern fits well in the standard picture~\cite{kofman:1997yn}, but in the spinodal region the resonant production appears to be more efficient than usual: upon spinodal zero-crossings the resonant production that normally shows (as it indeed does at later times also in our example) a period of anti-correlation, is here always positively correlated. While individual growth bursts are not enhanced, this positive correlation leads to particularly strong particle production.
%
%

%
\begin{figure}[t!]
\begin{center}
   \includegraphics[width=0.85\textwidth]{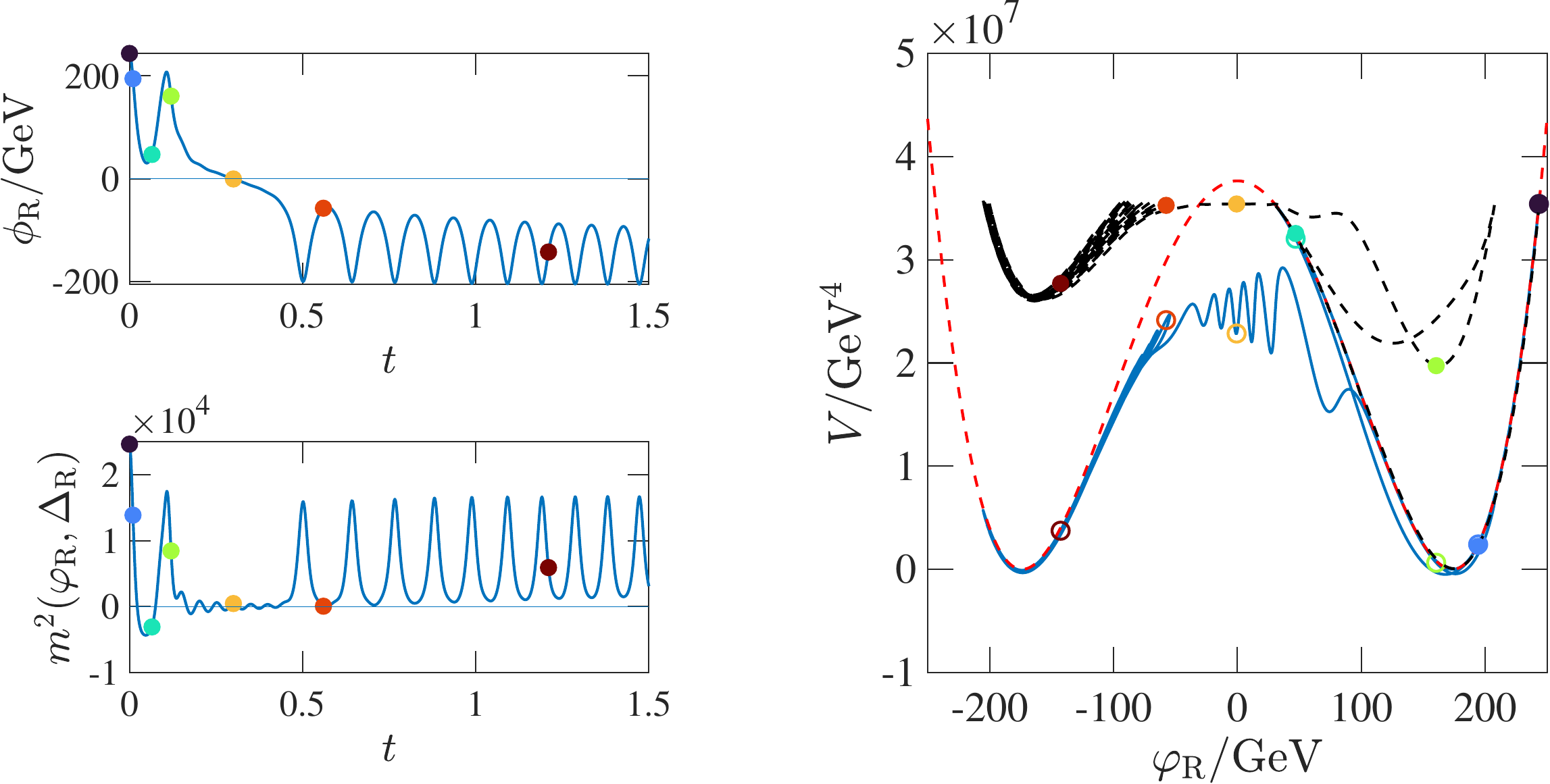} \hspace{1em}
\caption{The upper left panel shows the time evolution of $\phR$ (in units GeV) and the lower left panel that of the effective mass function $m^2(\phR,\GR)$ (in units GeV$^2$) in the case of a strong spinodal instability. In the right panel we show the time-evolution of the instantaneous effective potential~\cref{eq:V1PI} (dashed black line), embedded in a plot of the vacuum Hartree potential (dashed red line). The colored dots indicate select times at which the instantaneous potential was evaluated as indicated in the left panels. The solid blue line shows the instantaneous value of the non-equilibrium vacuum potential~\cref{eq:effpot}.}
\label{fig:wwo2pt}
\end{center}
\end{figure}
%

Because we did not include interactions in this run, the fluctuation band structure remains stable at all times. The system also remains highly coherent, as is evident both from the increase of the integrated coherence function and the stability of the heat plot of the coherence function shown in the right panels of figure~\cref{fig:figure5}.

%
\subsection{Strong spinodal instability}
\label{sec:out-of-eq-potential}
%

In the above analysis we made little reference to the effective potential. Indeed, the one-particle irreducible effective action is not a very useful quantity in an out-of-equilibrium setting and it can even be defined only \emph{after} the equations of motion have been solved. Even then one cannot define it universally, but only as a quantity evaluated locally in time. We will now study this question in the case of a very strong spinodal instability. To be specific, we still use the values $m_{\mathrm{R}} = 100$ GeV, $\lRidx{4} = 1$ and $\partial_t \varphi_{\rm R,in} = 0$, but we take  $\varphi_{\rm R,in} = 243.5$ GeV and include also friction. We assume that collisions drive the system to the vacuum state, i.e.\ we take $\delta \rho_{{n\bm k}}^{\rm eq} \equiv 0$, and we specify the coefficients to be $c_{1,2} = 0.6$ GeV.%
\footnote{Although we gave the friction terms only in a qualitative form, we can provide an estimate for the magnitude of the $c_i$-coefficients. From equations~\cref{eq:moments-fric} it is clear that $c_i$ have the dimensions of mass. The lowest order contribution to the collision integrals arises at the second order in coupling in the 2PI expansion. Hence the naïve scale of the coefficients $c_i$ is given by $\bigl(\frac{\lambda}{4\uppi}\bigr)^2m$, which for $\lRidx{4} = 1$ and $m_{\mathrm{R}} = 100$ GeV gives $c_i \simeq 0.6$ GeV.} 
In this case the initial potential energy of the field is lower than the peak of the vacuum potential at $\phR = 0$. This can be seen in the right panel of figure~\cref{fig:wwo2pt}, where we plot the Hartree-resummed vacuum potential (red dashed line) and indicate the initial field value by the black dot. 

\begin{figure}[t]
\begin{center}
   \includegraphics[width=0.95\textwidth]{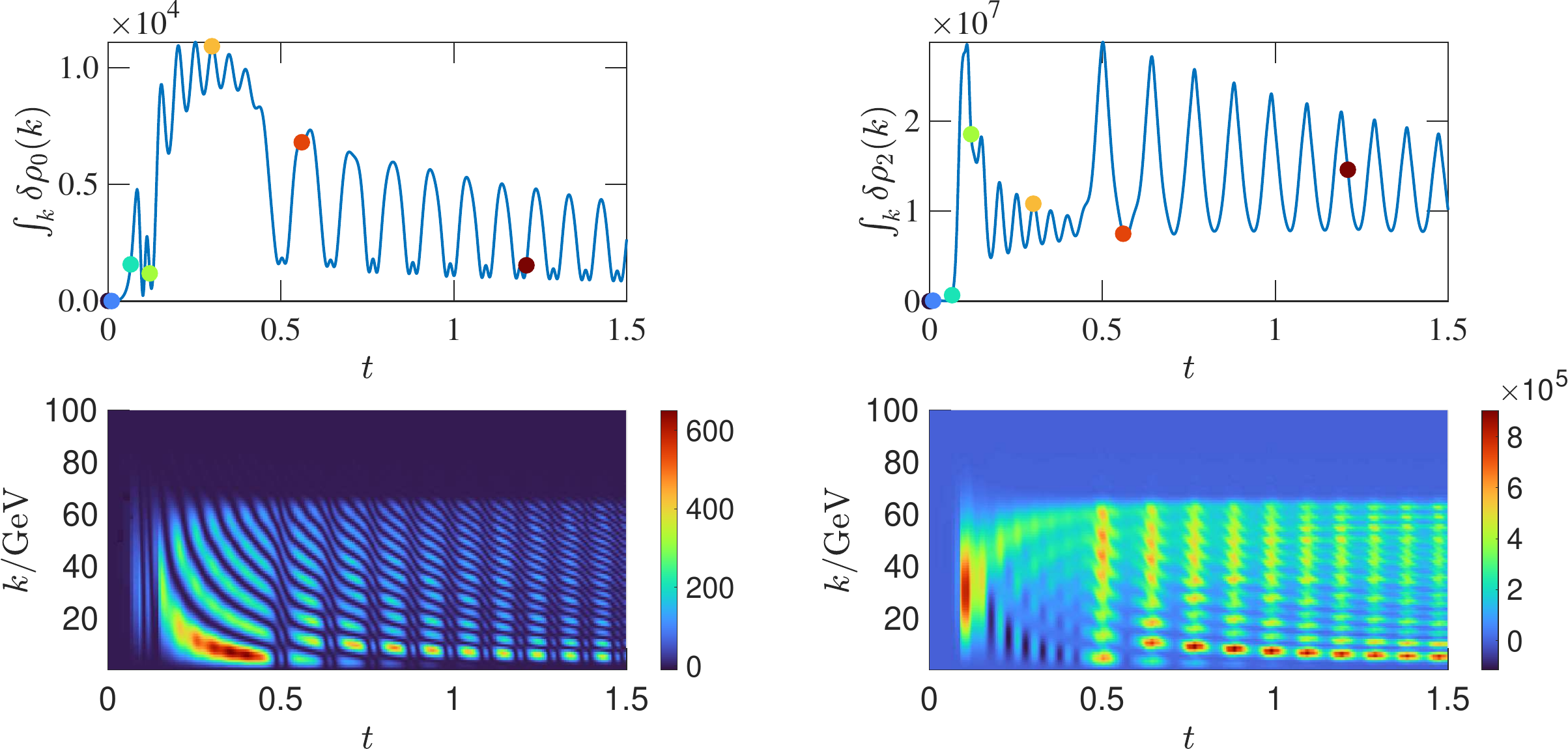}
\caption{The upper panels: shown are the integrated non-equilibrium fluctuations of the moment functions, $\int_{\bm{k}}\nolimits\delta\rho_{0,2\bm{k}}$. The colored dots have the same interpretation as in figure~\cref{fig:wwo2pt}. The lower panels: heat plots showing the momentum distributions $\frac{1}{2\uppi^2}{\bm k}^2\delta\rho_{n\bm{k}}$ corresponding to the upper panels. The left panels show the zeroth moment $n=0$ and the right panels the second moment $n=2$.}
\label{fig:kdistributions}
\end{center}
\end{figure}

Obviously, if the potential was held fixed, the field would simply oscillate around the positive minimum with a decaying amplitude. However, when backreaction is included, the picture changes dramatically. The actual field evolution is shown in the upper left panel of figure~\cref{fig:wwo2pt}. Curiously, the field stays around the positive minimum during only one oscillation cycle, after which it apparently passes through the potential barrier, spending a rather long time near the middle of the potential with the effective mass function close to zero. Of course what happens is that in the first passage of the field into the spinodal region, an explosive creation of fluctuations takes place. This is clearly demonstrated in figure~\cref{fig:kdistributions}, which shows the integrated fluctuations in the moment functions (upper panels) and the associated heat plots in the time-momentum plane (lower panels). These fluctuations absorb a large amount of entropy, which decreases the \emph{free energy} in the system and lowers the barrier between the minima allowing the field to pass to the negative side. The key issue is to not confuse the total internal energy of the system and the free energy, which may vary strongly depending on the entropy production. 

%
%

\paragraph{Non-equilibrium effective potentials.}
While the effective potential cannot be defined \emph{a priori}, it is illustrative to construct it \emph{a posteriori} as a time dependent potential that reproduces the equation of motion~\cref{eq:moments-phi} at all times. This potential can be constructed as the definite integral
\begin{equation}
V_{\rm 1PI}(t;\phR) \equiv \int_{t_{\mathrm{in}}}^t \biggl[ -\frac{1}{3}\lambda^\idx{2}_{\rm R}\varphi_{\rm R}^3 + m^2(\phR,\GR)\phR\biggr](\partial_{\tilde{t}} \phR)\mathrm{d}\tilde{t},
\label{eq:V1PI}
\end{equation}
where $\phR$ and $\GR$ are the solutions of the equations of motion. We show this potential as the dashed black line in figure~\cref{fig:wwo2pt}. After the crossing to the negative side, the shape of the potential function settles and the field oscillates around the negative minimum with a decaying amplitude. We stress that $V_{\rm 1PI}$ is only useful for the visualization and interpretation of results and there is no unique definition of the effective potential in the non-equilibrium case. 

As was already mentioned in section~\cref{sec:effpot}, in any finite truncation the renormalized 2PI vacuum becomes dependent on the IR-physics. Another interesting potential\footnote{In reference~\cite{PhysRevD.65.065019} yet another dynamical potential was defined as the difference between the total energy of the system and the kinetic energy of the classical field.} function then is the equivalent of the vacuum Hartree potential in the presence of fluctuations. This potential is defined as
\begin{equation}
V_{\rm H\G}(\phR,\GR) \equiv V_{\rm H}(\phR,\GR) - \frac{1}{2}m^2(\phR,\GR)\int_{\bm k}\nolimits \delta \rho_{0\bm{k}},
\label{eq:effpot}
\end{equation}
where $V_{\rm H}(\phR,\GR)$ is the 2PI vacuum potential~\cref{2PI-effective-potential-final} evaluated replacing the vacuum mass function $\overbar m^2(\phR)$ with the general mass function $m^2(\phR,\GR)$. Note that the integral term over the fluctuations of the zeroth moment is a part of the vacuum Hartree potential, similarly to the case with the thermal potential~\cref{eq:finite-T-pot}. The potential~\cref{eq:effpot} is shown with the blue solid line in the right panel of figure~\cref{fig:wwo2pt}. It represents changes in the 2PI Hartree vacuum energy including the backreaction effects, and like the instantaneous $V_{\rm 1PI}$-potential, its barrier around $\phR=0$ is temporarily lowered by the backreaction. This example demonstrates that the final stages of a phase transition may involve very complicated quantum dynamics, where classical expectations and constraints do not hold. 

We conclude this subsection by stressing on the difference of the fluctuation spectra in the present case, shown in the lower panels of figure~\cref{fig:kdistributions}, and in the parametric resonance case shown in figure~\cref{fig:figure5}. Even though we used the same mass and coupling parameters, essentially all fluctuations are here created by the spinodal instability. Indeed, they occupy a region in the phase space which is consistent with the instability constraint~\cref{eq:spinodal_modes}, continues all the way to zero momentum and lies entirely below the parametric resonance band.

%
\begin{figure}[t]
\begin{center}
\includegraphics[width=0.85\textwidth]{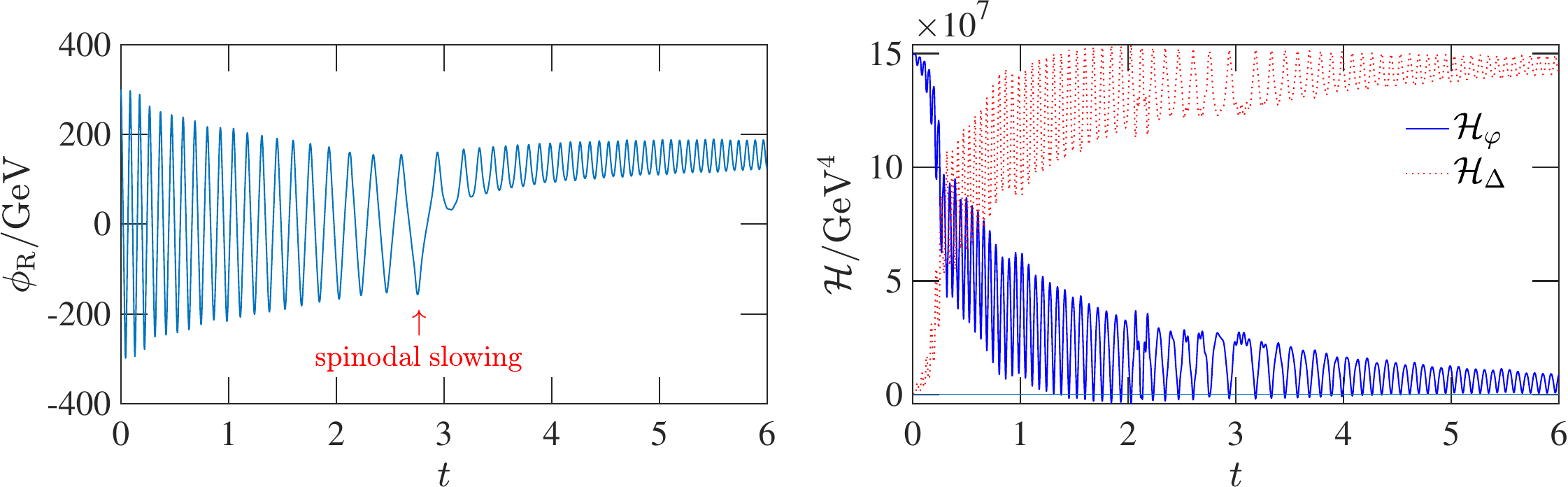}\hspace{1em}
\end{center}
\caption{Shown is the time-evolution of the classical field (left panel) and that of the total energy in the fluctuations and the classical field (right panel). ${\mathcal H}_\varphi(t)$ is the energy in the classical field and ${\mathcal H}_\G(t)$ is the energy in the fluctuations. The physical parameters and the specific form of the collision integrals used in this run are described in the text.}
\label{fig:figure8}
\end{figure}
%

\subsection{Self-thermalization}
\label{sec:self-thermalization}
%

%
\begin{figure}[t]
\begin{center}
\includegraphics[width=0.95\textwidth]{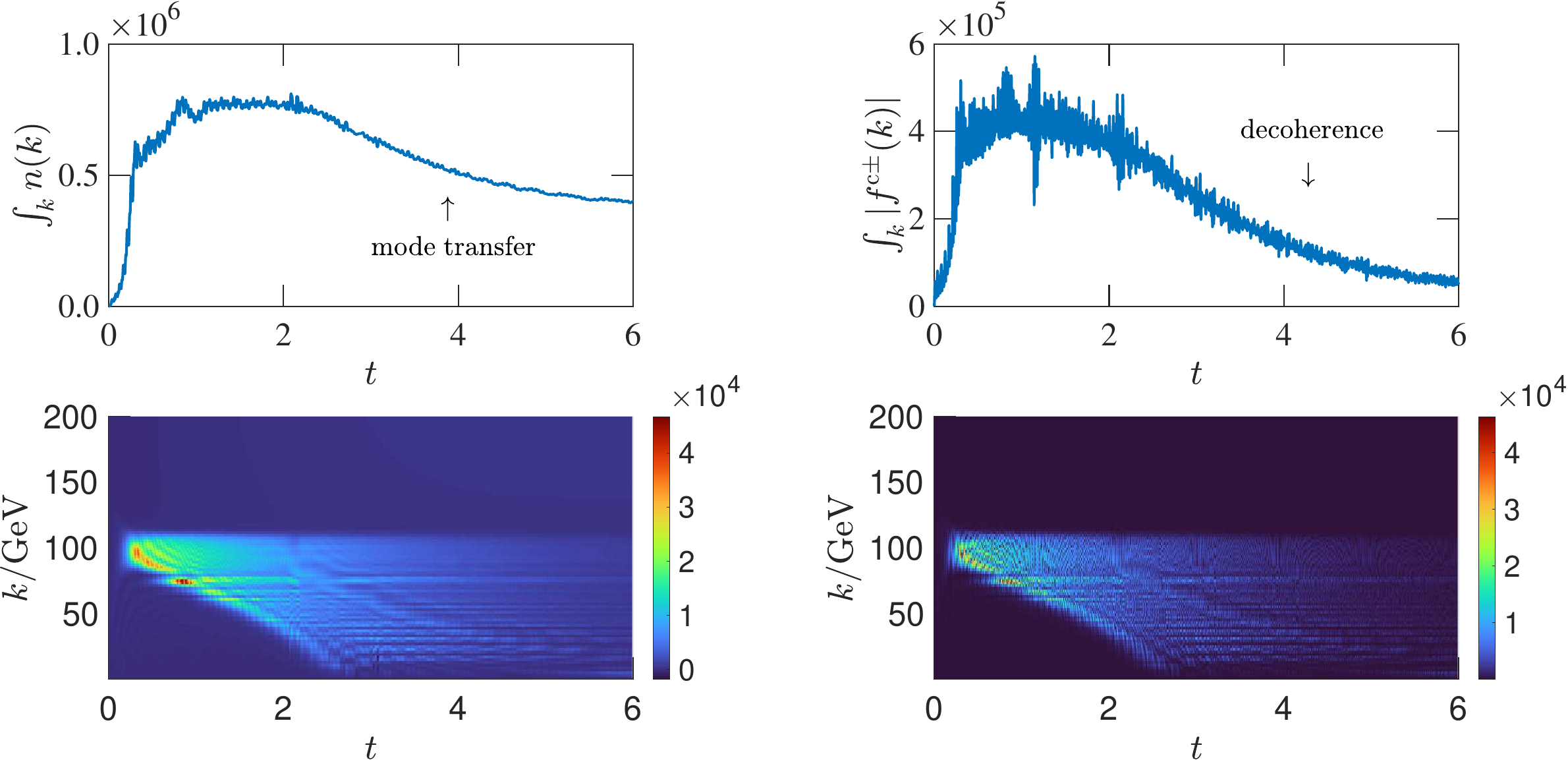}
\end{center}
\caption{Shown are the evolution of the number density (left) and the modulus of the coherence functions (right). In the upper panels the quantities are integrated over momentum. We used the same parameters as in figure~\cref{fig:figure5}, except for non-zero friction coefficients $c_i=0.6$ GeV in the collision integrals with thermal equilibrium solutions.}
\label{fig:figure9}
\end{figure}
%

As our final example we study thermalization of the scalar field energy in a self-interacting system. We use the same physical parameters and initial conditions as in section~\cref{sec:parmetric-resoanance} but include collision terms with the friction coefficients $c_{0,1}=0.6$ GeV, and assume that the collisions drive the system to thermal equilibrium, i.e.\ we take $\delta \rho_{n\bm{k}}^{\rm eq} \equiv \delta \rho_{n\bm{k}}^{\rm th}$. With rigorously computed collision terms the thermal state would emerge automatically as an attractor solution, but in our phenomenological approach we need to give a definition for the instantaneous temperature. In thermal equilibrium a general moment can be written as 
\begin{equation}
\rho_{n{\bm k}}^{\mathrm{th}} = \frac{1}{2}\,\omega_{\bm k}^{n-1}
  \Bigl[ n_\mathrm{BE}(\omega_{\bm k}) + (-1)^n \bigl( 1 + n_\mathrm{BE}(\omega_{\bm k}) \bigr) \Bigr ],
\end{equation}
where $n_{\mathrm{BE}}(k_0) = (\mathrm{e}^{k_0/T} - 1)^{-1}$ is the Bose--Einstein distribution function. In particular
\begin{equation}
  \delta\rho_{0{\bm k}}^{\mathrm{th}} = \frac{1}{\omega_{\bm k}}n_\mathrm{BE}(\omega_{\bm k})
\quad \mathrm{and} \quad
  \delta\rho_{2{\bm k}}^{\mathrm{th}} = \omega_{\bm k}n_\mathrm{BE}(\omega_{\bm k}).
\end{equation}
while $\delta \rho_{1\bm{k}}^{\rm th} = 0$. We define the equivalent temperature $T = T(t)$ by requiring that the thermal state has the same energy as what is stored in the fluctuations:
\begin{equation}
{\mathcal H}_\Delta(t) \equiv \int_{\bm k}\nolimits \delta \rho_{2\bm{k}}(t) \equiv \int_{\bm k} \nolimits \omega_{\bm k} n_{\rm BE}(\omega_{\bm k}).
\label{eq:equivalent-temperature}
\end{equation}
In all these equations $\omega_{\bm k}^2 = {\bm k}^2 + m^2(\phR,\GR)$ is a function of time. The energy stored in the classical field is
\begin{equation}
{\mathcal H}_\varphi(t) \equiv \frac{1}{2}\bigl(\partial_t\varphi_{\rm R}(t)\bigr)^2 + V_{\rm H\G}(\phR(t),\GR(t)).
\label{eq:Hvarphi}
\end{equation}
With our definitions of the temperature and the collision integrals the total energy ${\mathcal H} = \mathcal{H}_\varphi + \mathcal{H}_\Delta$ should be conserved, and we checked that this is indeed the case to a high accuracy in our calculations. For more details on this, and on the numerical setup in general, see appendix~\cref{sec:impl}.

\paragraph{Spinodal slowing.}

In the left panel of figure~\cref{fig:figure8} we show the evolution of the classical field $\phR$. Initially $\phR$ evolves as in the collisionless case, oscillating with a nearly constant frequency and a large amplitude, but around $t\sim 2$ the frequency starts to decrease until it reaches a minimum around $t\sim 3$. After this the field gets trapped around the positive minimum while the oscillation frequency increases again. This \emph{spinodal slowing} effect was already seen in connection with the barrier crossing in section~\cref{sec:out-of-eq-potential}. The bearing of the spinodal modes is revealed in the inset in the left panel of figure~\cref{fig:temperature-and-w}, which shows that the effective mass term $m^2(\phR,\GR)$ repeatedly becomes negative in this region. In the right panel of figure~\cref{fig:figure8} we show the energy components ${\mathcal H}_\varphi$ and $\mathcal{H}_\Delta$. Initially all energy is stored in the classical field, but the fraction of energy in the fluctuations increases until the system is {\em reheated}, with almost all of the energy contained in the fluctuations.

%
\begin{figure}[t]
\begin{center}
\includegraphics[width=0.45\textwidth]{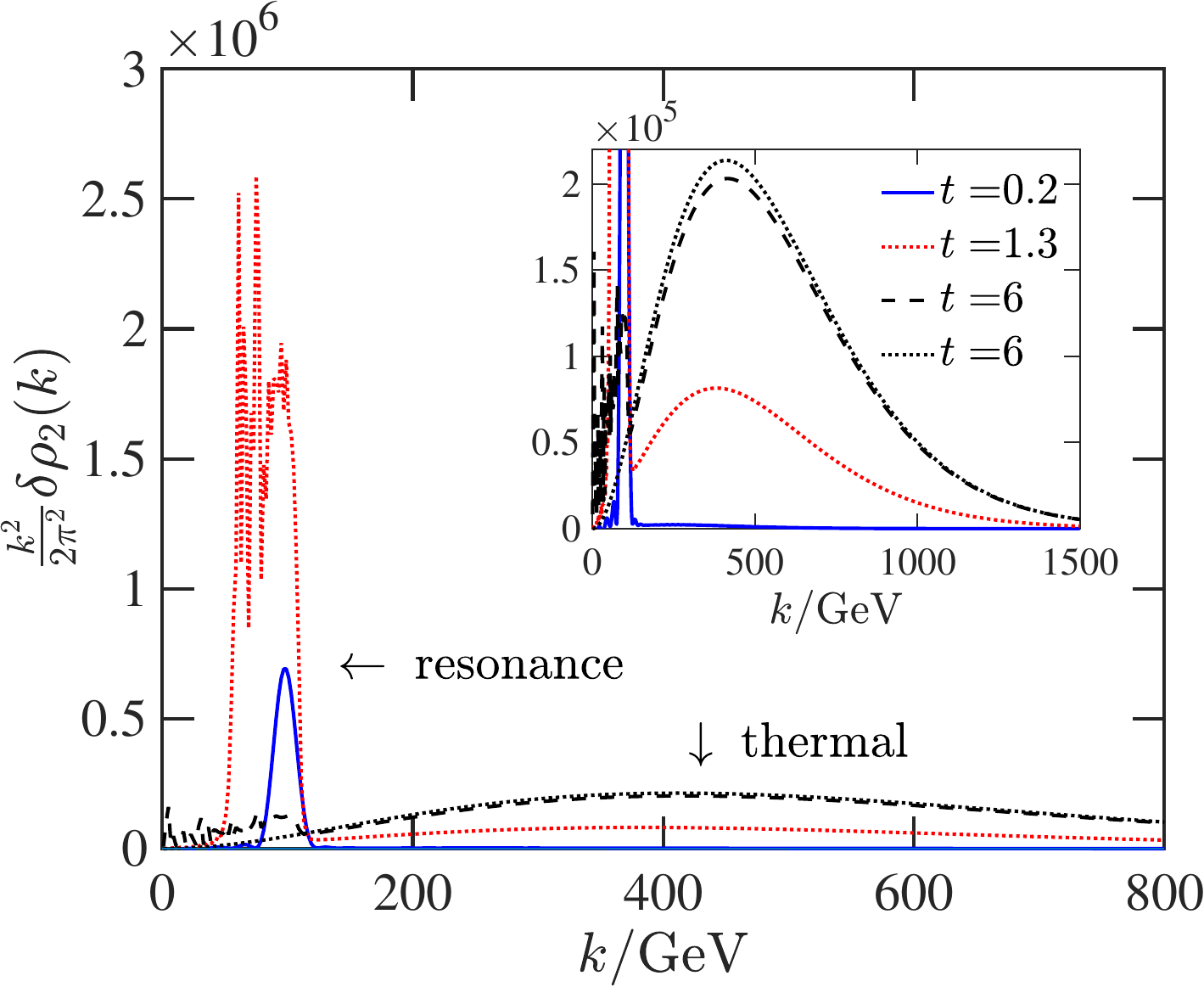} \hspace{1em}
\includegraphics[width=0.45\textwidth]{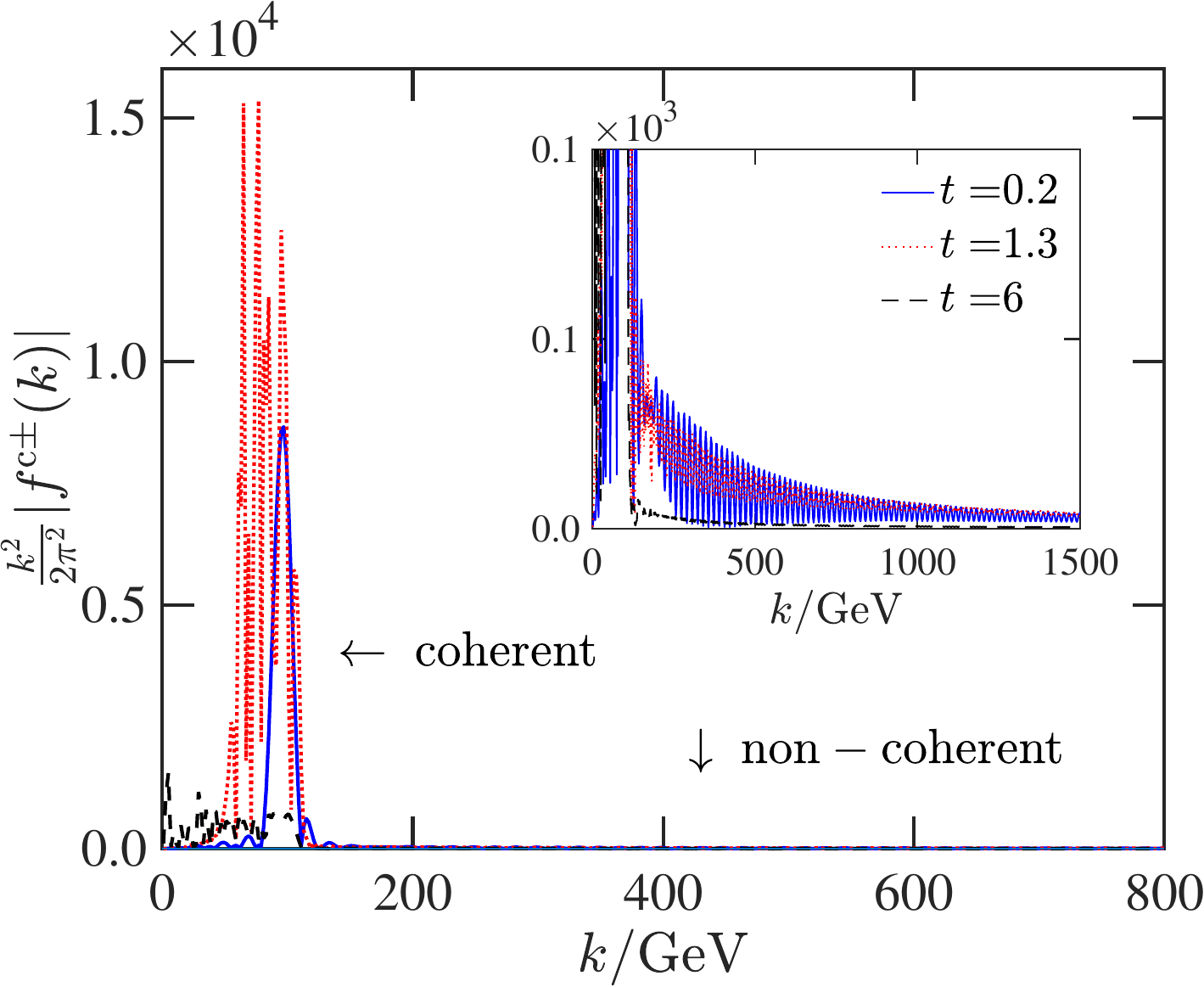}
\end{center}
\caption{Shown are the momentum distributions $\frac{\bm{k}^2}{2\uppi^2}\delta\rho_{2\bm k}$ (left) and $\frac{\bm{k}^2}{2\uppi^2}\bigl|f^{\rm c\pm}_{\bm k}\bigr|$ (right) for three different times: $t=0.2$ (solid blue lines) $t=1.3$ (red dotted lines) and $t=6$ (black dashed lines). Also shown in the left plot is the weighted thermal distribution $\frac{\bm{k}^2}{2\uppi^2}\omega_{\bm k}n_{\rm BE}(\omega_{\bm k})$ for the equivalent temperature $T(t=6)=144.9$ GeV (black dotted line).}
\label{fig:profiles_with_insets}
\end{figure}
%

\paragraph{Mode transfer and decoherence.}

In figure~\cref{fig:figure9} we again show the evolution of the number density and coherence functions, including both the integrated quantities and the time-momentum heat plots. There are striking, but expected differences between these plots and the corresponding non-interacting results shown in figure~\cref{fig:figure5}. First, the number density stops growing already at $t \sim 1$ and eventually starts to decrease for $t\gtrsim 2$. As is seen from figure~\cref{fig:figure8}, fluctuations dominate the total energy already for $t \gtrsim 1$, and the subsequent decrease of particle number results from a transfer of modes to higher energies. Thermalization process should also lead to decoherence, and this is indeed clearly visible in the upper right panel of figure~\cref{fig:figure9}, which shows the integrated function $\bigl|f^{c\pm}_{\bm k}\bigr|$. From the heat plots we see that particle production gets progressively less efficient and moves to smaller frequencies, as less and less energy is left in the classical field. From the heat plot in the lower right panel we see that coherence is erased throughout the phase space at late times.

\paragraph{Thermalization.}

In figure~\cref{fig:profiles_with_insets} we show the $|\bm k|$-distributions of $\delta \rho_{2\bm{k}}$ (left panel) and the coherence function $\bigl|f^{c\pm}_{\bm{k}}\bigr|$ (right panel) weighted by the phase space factor, for selected times during the evolution. At a relatively early time $t=0.2$ the distributions shown in solid blue still display a clear parametric resonance band structure. At a later time $t=1.3$ (red dotted lines) the resonant spectrum is already much more complex, apparently with contributions from many narrow bands. Also a significant mode-transfer to the thermal region has already taken place. Indeed, from the main plot in the left panel of figure~\cref{fig:temperature-and-w} we see that the equivalent temperature at $t=1.3$ is roughly 140 GeV, and as the field is relatively light, $\langle m^2_{\rm eff} \rangle^{1/2}/T \lesssim 1$ with $\langle m^2_{\rm eff} \rangle$ being the local average of the oscillating effective mass function, the expected maximum of the thermal spectrum is located at $\langle |{\bm k}|\rangle \approx 3T \approx 400$ GeV. At the end of the simulation, $t=6$ (black dashed curve), the system has essentially thermalized. Almost all energy is in the fluctuations and very little particle production activity remains. The particle number in the resonance bands is small and the coherence is almost vanishing everywhere and in particular in the thermal region. Also the fluctuations in the equivalent temperature have but a small residual amplitude left. For the final time we also plotted (black dotted line in the left panel of figure~\cref{fig:profiles_with_insets}) the equivalent thermal spectrum $\frac{\bm{k}^2}{2\uppi^2}\omega_{\bm k}n_{\rm BE}(\omega_{\bm k})$ with $T = 144.9$ GeV, corresponding to the equivalent temperature at $t=6$. The close agreement between the actual and thermal distributions shows that the system has indeed thermalized to a very high accuracy.

%
\begin{figure}[t]
\begin{center}
\includegraphics[width=0.45\textwidth]{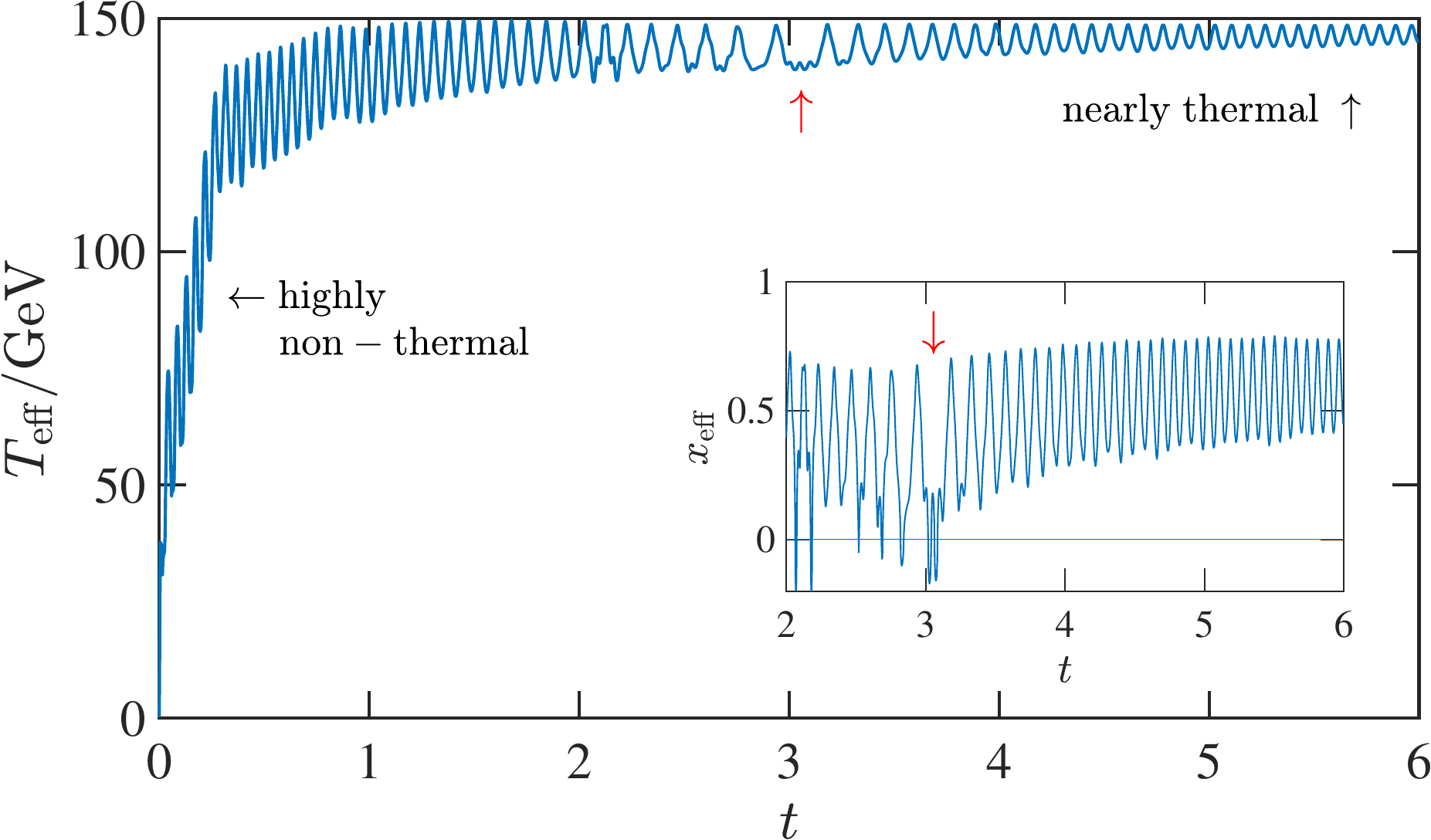}
\includegraphics[width=0.45\textwidth]{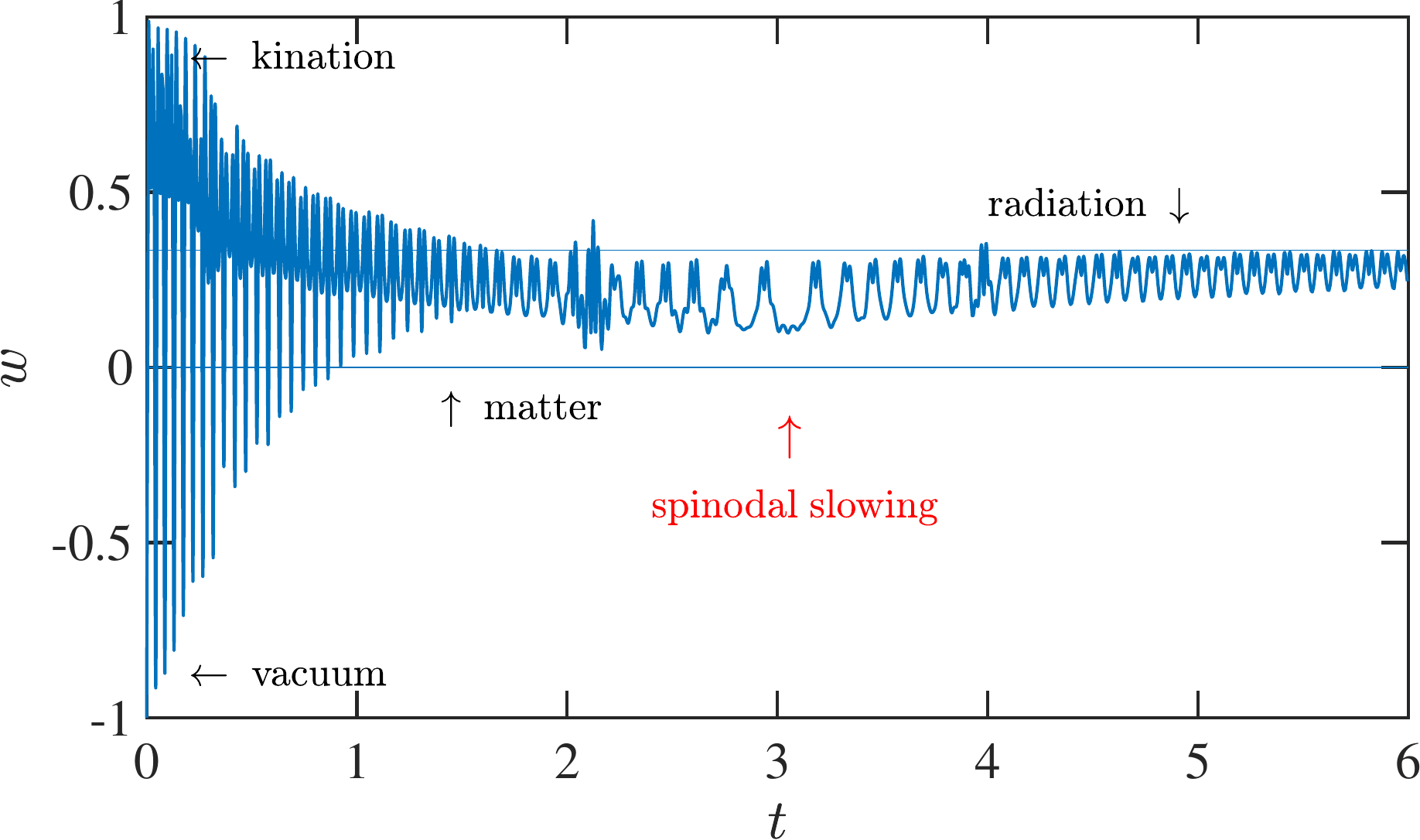}
\end{center}
\caption{In the left panel we show the equivalent temperature defined through equation~\cref{eq:equivalent-temperature} as a function of time. The inset shows the parameter $x_{\rm eff} \equiv {\rm sgn}\bigl(m^2_{\rm eff}\bigr)\bigl|m^2_{\rm eff}\bigr|^{1/2}/T$. In the right panel we show the EOS-parameter of the system defined in equation~\cref{eq:EOS}. The black arrows indicate the limiting cases of vacuum ($w=-1$) and kinetic ($w=1$) energy dominance as well as matter ($w=0$) and radiation ($w=1/3$) EOS's, shown by horizontal lines. In all graphs shown the red arrow points the region of maximal spinodal slowing.}
\label{fig:temperature-and-w}
\end{figure}
%

\paragraph{Equation of state.}

Let us finally study the evolution of the equation of state (EOS) in the system. The EOS-parameter is defined as
\begin{equation}
w \equiv \frac{\mathcal P}{\mathcal H},
\label{eq:EOS}
\end{equation}
where ${\mathcal H} = {\mathcal H}_\varphi + {\mathcal H}_\G$ is the total energy and the total pressure  ${\mathcal P} = {\mathcal P}_\varphi + {\mathcal P}_\G$  is similarly the sum of the pressures in the classical field and in the fluctuations. The former is given by
\begin{equation}
{\mathcal P}_\varphi = \frac{1}{2}(\partial_t\phR)^2 - V_{\rm H\G}(\phR,\GR),
\label{eq:pressure-in-fields}
\end{equation}
where $V_{\rm H\G}$ was defined in~\cref{eq:effpot}. The pressure contained in the fluctuations can be computed as the spatial component of the energy-momentum tensor~\cite{Herranen:2008di}, and it can be written in terms of the moment functions as follows:
\begin{equation}
{\mathcal P}_\G(\phR,\GR) =  \int_{\bm k} \nolimits \biggl[ \delta \rho_{2\bm k}(t) + 
\biggl( \frac{1}{3}{\bm k}^2 - \omega_{\bm k}^2 \biggr) \delta \rho_{0\bm k}(t) \biggr].
\label{eq:PcQPA}
\end{equation}
It is easy to see that in the thermal limit~\cref{eq:PcQPA} reduces to the negative of the thermal part of the effective potential in the Hartree approximation: ${\mathcal P}_\G =- T^4 {\mathcal J}\bigl( \overbar m^2_T/T^2\bigr)$.

We plot the EOS-parameter $w$ in the right panel of figure~\cref{fig:temperature-and-w}. The EOS-parameter starts from $w=-1$ and initially oscillates between $w=-1$, corresponding to total vacuum energy dominance, and $w=1$, corresponding to kinetic energy dominance (kination) in the classical field sector. However, as the energy is moved out from the field and the system thermalizes, the EOS-parameter moves to the band $0 < w < 1/3$ corresponding to normal matter. From the inset of the left panel we see that the average value $\langle |x_{\rm eff}|\rangle = \langle |m^2_{\rm eff}|^{1/2}/T\rangle \approx 0.6$ at late times. This indicates that the reheated thermal plasma is almost relativistic and indeed, the EOS-parameter is asymptoting close to $w=1/3$ at late times. (In a purely thermal plasma with $x_{\rm eff}=0.6$ one would get $w\approx 0.315$.) The periodic deviation below this value seen in figure~\cref{fig:temperature-and-w} is due to the field contributions to energy and pressure.

%
\section{Conclusions}
\label{sec:conc}
%

We have studied the non-equilibrium evolution of a system consisting of a classical scalar field coupled to the two-point function describing quantum fluctuations. We derived renormalized evolution equations for the system using 2PI methods in the Hartree approximation. We derived the effective potential for this system in vacuum and in thermal equilibrium and compared the latter with the known one-loop-resummed effective potentials. We showed that the Parwani-resummed thermal potential~\cite{Parwani:1991gq} is closest in spirit to the Hartree-resummed effective potential. We showed that in a non-equilibrium situation the 2PI method, in any finite truncation, leads to an effective vacuum potential (the vacuum state) that depends on the infrared physics. Indeed, even though the renormalization procedure provides unique and constant counterterms, the split of the system into divergent and non-divergent parts depends on the IR-physics.

We wrote our renormalized evolution equations as a set of coupled moment-equations for the correlation function and a field equation for the one-point function in the mixed representation and included phenomenological collision integrals describing friction. We used this system to study the non-perturbative particle production and spinodal instability at the end of phase transitions. We found out that quantum backreaction can have significant effects on the evolution of the system and addressed the problems in trying to define any practical effective potential for such dynamical systems. In particular we were able to follow the full thermal history of a self-interacting system starting from a cold initial state where all energy in the system was stored in the classical potential, until the end when the system was reheated and thermalized and the field stayed at the minimum of the thermal (Hartree) effective potential.

In this work we assumed that the quantum system lived in the Minkowski space-time. Generalization to an expanding FRLW space-time is straightforward by a simple transform to conformal coordinates~\cite{Jukkala:2021cys}. Moreover, in many realistic systems the time scales involved in the phase transition are much faster than the Hubble expansion. In those cases our results are representative of the physics as such. Also, we used only a phenomenological form for the collision integrals. It would be interesting to derive more realistic collision terms using the methods developed in~\cite{Herranen:2010mh,Fidler:2011yq}. Also it would be interesting to couple the scalar field also to other quantum fields. This should be straightforward by combining the current results with the quantum transport equations for fermions developed in~\cite{Jukkala:2019slc}. In this way one should be able to study reheating at the end of inflation in a realistic setup.

%
\section*{Acknowledgements}
\label{sec:ack}
%

This work was supported by the Academy of Finland grant 318319. OK was in addition supported by a grant from the Magnus Ehrnrooth Foundation. We wish to thank Alexandre Alvarez, Amitayus Banik, Haye Hinrichsen, Sami Nurmi, Werner Porod and Anna Tokareva for discussions and comments on the manuscript.

%
\appendixtitleon
\appendixtitletocon
\begin{appendices}
%
%
\section{Numerical implementation}
\label{sec:impl}
%

In this appendix we discuss some technical points that are relevant for an accurate and efficient solution of the evolution equations. The first one concerns identifying a conserved quantity in the non-interacting limit. The equations rewritten using this variable are much more stable than the original equations. The second point concerns discretization. In a naïve binning of the momentum variable, the discrete integral of the vacuum term in equation~\cref{eq:meff-gap-equation} is badly behaved numerically near the edges of the spinodal regions. This problem can be avoided by a more careful definition of the binned variables. Finally, we show how our numerical setup conserves the total energy of the solved system to a high accuracy with the self-thermalizing system as a case study.

\paragraph{Stabilized equations.} 

It was noted already in reference~\cite{Herranen:2008di} that the moment equations~\cref{eq:moments-rho0,eq:moments-rho1,eq:moments-rho2} can be written in a form that is more resistant to numerical instabilities, using the variable
\begin{equation}
\label{eq:X}
X_{\bm k} \equiv 2\rho_{0\bm{k}} \rho_{2\bm{k}} - \omega_{\bm{k}}^2(t) \rho_{0\bm{k}}^2 - \frac{1}{4}(\partial_t \rho_{0\bm{k}})^2.
\end{equation}
Indeed, if we multiply~\cref{eq:moments-rho0} by $2\partial_t \rho_{0\bm{k}}$ and~\cref{eq:moments-rho2} by $2\rho_{0\bm{k}}$ and subtract the resulting equations, we can show that $X_{\bm k}$ is conserved in the collisionless limit: $\partial_t X_{\bm k} = 0$. With non-vanishing friction terms $X_{\bm k}$ is no longer conserved, but the derivation with equations~\cref{eq:moments-fric} including friction proceeds analogously, and one finds:
\begin{subequations}
\label{eq:finalesomstab}
\begin{align}
\frac{1}{4}\partial_t^2\rho_{0\bm{k}} - \rho_{2\bm{k}} + \omega_{\bm{k}}^2(t) \rho_{0\bm{k}}
&= -c_1\partial_t\rho_{0\bm{k}},  \\
\partial_t \rho_{1\bm{k}} &= -c_2\bigl(\delta\rho_{1\bm{k}} - \delta\rho_{1\bm{k}}^{\mathrm{eq}}\bigr), \\[.3em]
\partial_{t}X_{\bm k} 
&= 2{c_1}\bigl(\partial_t \rho_{0{\bm k}}\bigr)^2 -2{c_2}\rho_{0{\bm k}}\bigl(\delta\rho_{2{\bm k}} - \delta\rho_{2{\bm k}}^{\mathrm{eq}}\bigr).
\end{align}
\end{subequations}
We have thus replaced $\rho_{2{\bm k}}$ by $X_{\bm k}$ as a dynamical variable. We will use~\cref{eq:X} to set the initial condition for $X_{\bm k}$  in terms of the initial values for \(\rho_{0\bm k}\), \(\partial_t \rho_{0\bm k}\) and $\rho_{2\bm k}$, and at any point during and at the end of the calculation we can compute $\rho_{2{\bm k}}$ from $X_{\bm k}$ using the inverse relation
\begin{equation}
\rho_{2\bm{k}} = \frac{1}{2\rho_{0\bm{k}}}\biggl[X_{\bm k} + \frac{1}{4}(\partial_t \rho_{0\bm{k}})^2 + \omega_{\bm{k}}^2(t)\rho_{0\bm{k}}^2 \biggr].
\end{equation}

\paragraph{Coarse-grained binning.} 

Whenever the effective mass term is negative there is a momentum for which $m^2(\phR,\GR) = -{\bm k}^2$ and at this point the zeroth momentum vacuum function $\rho_{0\bm{k}}^{\mathrm{vac}}=\Theta_{\bm k}/(2\omega_{\bm k})$ diverges. This is a mild, integrable singularity that does not affect the continuum limit, but it can cause overflows and numerical inaccuracy in a system with a finite discretization. This problem can be avoided by a careful choice of binned variables for the vacuum distribution. That is, we replace the vacuum distribution by a coarse-grained distribution defined by an integration over each momentum bin $q \in [q_i,q_{i+1}]$:
\begin{equation}
\frac{1}{2\omega_{q_{ci}}} \rightarrow \frac{1}{2q_{ci}^2\Delta q_i} \big[ i_0(q_{i+1}) - i_0(q_i) \big],
\label{eq:replacement}
\end{equation}
where $q_{ci} \equiv \frac{1}{2}(q_i+q_{i+1})$, $\Delta q_i \equiv q_{i+1} - q_i$  and
\begin{equation}
i_0(q) \equiv \frac{1}{2}  \biggl[ q\omega_{q} - m^2{\rm artanh}\biggl(\frac{q}{\omega_{q}}\biggr) \biggr].
\end{equation}
When the bin width goes to zero, the replacement~\cref{eq:replacement} does not make any difference. However, for a finite discretization it avoids the singularity that would occur in the spinodal region when the effective mass function coincides with one of the bin-momenta squared, $m^2(\phR,\GR)= - q_{ci}^2$.

%
\begin{figure}[t]
\begin{center}
   \includegraphics[width=0.8\textwidth]{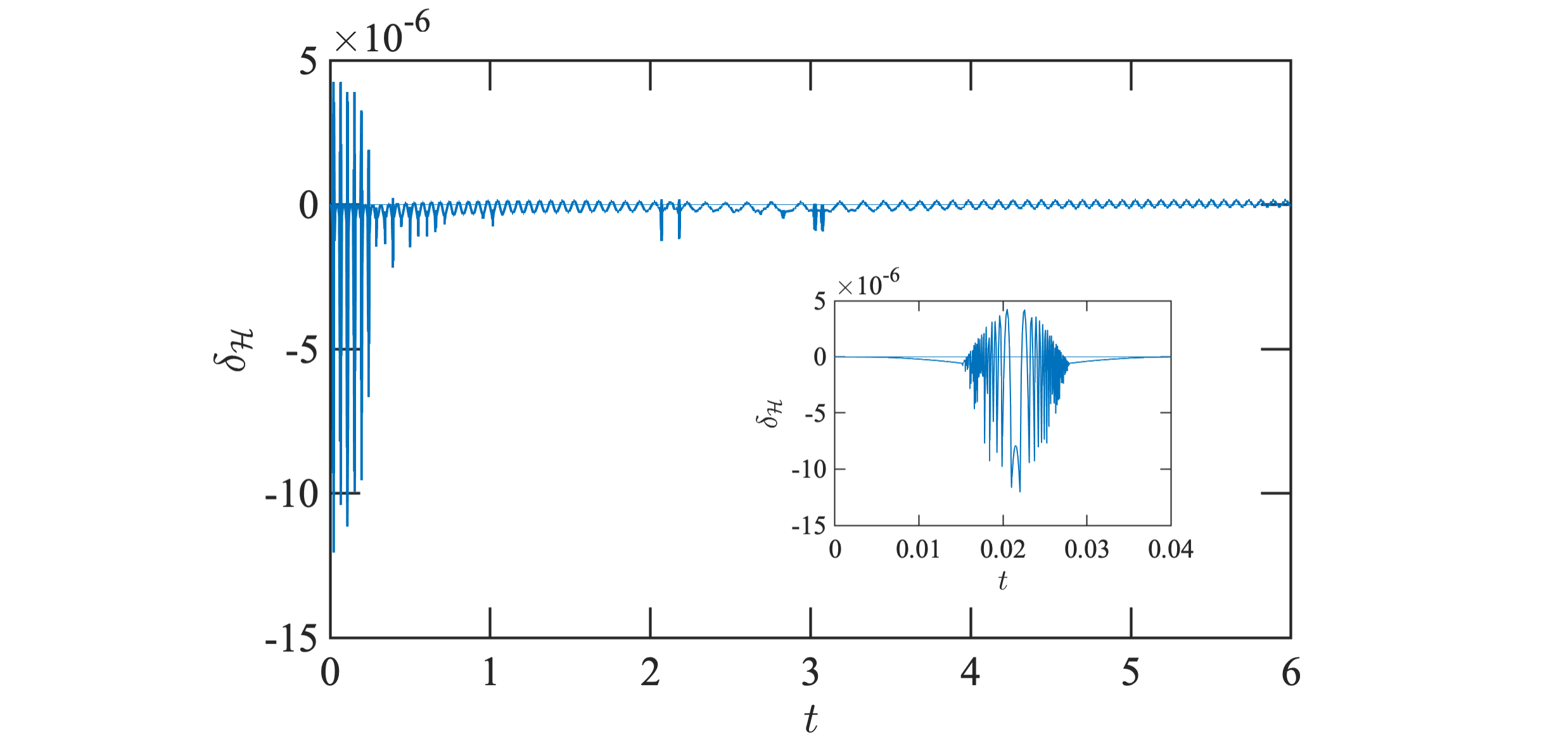}\hspace{2em}
\caption{Shown is the relative change in energy $\delta_{\mathcal{H}} = \mathcal{H}/\mathcal{H}_0-1$ during calculation in the self-thermalization case studied in section~\cref{sec:self-thermalization}. Inset shows a close-up on the first spinodal instability region.}
\label{fig:error_estmates}
\end{center}
\end{figure}
%

\paragraph{Energy conservation.} 

In figure~\cref{fig:error_estmates} we show the relative change in the total energy $\delta_{\mathcal H} \equiv {\mathcal H}/{\mathcal H}_0 - 1$ in the example we studied in section~\cref{sec:self-thermalization}. The total energy is ${\mathcal H} = {\mathcal H}_\varphi + {\mathcal H}_\G$, where  partial energies in the fluctuations ${\mathcal H}_\Delta$ and in the classical field ${\mathcal H}_\varphi$ were defined in equations~\cref{eq:equivalent-temperature,eq:Hvarphi}. In this example the total energy should be conserved, and this is indeed true to a very high accuracy. In this run we used a discretized momentum $|\bm{k}| \in [0, 2000]$ GeV with 1000 grid points. As can be seen in the figure, the error is essentially negligible between the spinodal regions. Within the spinodal regions there is some residual noise at early times. This arises from the integrable singularity near $m^2(\phR,\GR)=0$, even with the coarse grained binning, but even this error is small and can be further reduced by reducing the bin width.  We conclude that numerical errors are well under control in our calculations. 

\end{appendices}

%
\bibliography{main.bib}
%

%
\end{document}